\documentclass[nobibnotes,nofootinbib,superscriptaddress]{revtex4}
\usepackage{amsmath,amssymb,bm}
\usepackage{graphicx}
\usepackage{pstricks,pst-node,pst-text,pst-3d}
\usepackage{datetime}
\usepackage{verbatim}   
\usepackage{color}      
\usepackage{subfigure}  
\usepackage{hyperref}   

\def\xb{{\bf x}}    
\def\yb{{\bf y}}    
  
\def\xpb{{\bf x}_1} 
\def\xdb{{\bf x}_2} 
\def\xzb{{\bf x}_0}

\def\rb{{\bf r}}    
\def\bb{{\bf b}}

\def\asb{\overline{\alpha}_s}
\newcommand{\beeq}{\begin{eqnarray}}
\newcommand{\eeeq}{\end{eqnarray}}
\newcommand{\be}{\begin{equation}}
\newcommand{\ee}{\end{equation}}
\newcommand{\bea}{\begin{array}}
\newcommand{\eea}{\end{array}}

\def\wb{{\bf w}}

\begin{document}
\title{\bf Numerical solution of the nonlinear evolution equation at small $x$ with impact parameter and beyond the LL approximation}

\author{Jeffrey Berger}\email{jxb1024@psu.edu}
\affiliation{Penn State University, Physics Department, University Park, PA 16802, USA}

\author{Anna M. Sta\'sto}\email{astasto@phys.psu.edu}
\affiliation{Penn State University, Physics Department, University Park, PA 16802, USA}
\affiliation{RIKEN Center, Brookhaven National Laboratory, Upton, NY 11973, USA}
\affiliation{Institute of Nuclear Physics Polish Academy of Sciences, Cracow, Poland}

\begin{abstract}
Nonlinear evolution equation  at small $x$ with impact parameter dependence is analyzed numerically. Saturation scales and the radius of expansion in  impact parameter are extracted as functions of rapidity. Running coupling is included in this evolution, and it is found that the solution is sensitive to the infrared regularization.  Kinematical effects beyond leading logarithmic approximation are taken partially into account by modifying the kernel which includes the rapidity dependent cuts.  While the local nonlinear evolution is not very sensitive to these effects,
the kinematical constraints cannot be neglected in the evolution with impact parameter. 
 \end{abstract}

\maketitle

\section{Introduction}

High energy limit in Quantum Chromodynamics is one of the most intriguing problems in strong interaction physics.  Since the Large Hadron Collider  has recently opened a new kinematic regime, it is of vital importance to understand how one can calculate the cross section at high energies from basic principles in QCD.  The BFKL equation obtained in the Regge limit of QCD   \cite{Fadin:1975cb,Kuraev:1977fs,Balitsky:1978ic,Lipatov:1985uk} predicted fast growth of the cross section with the energy, due to the exchange of the hard Pomeron. This behavior is  however  known to violate the unitarity bound of the scattering amplitude. Higher order corrections to the BFKL \cite{Fadin:1996nw,Fadin:1996zv,Fadin:1998py,Kotsky:1998ug,Camici:1996st,Camici:1996fr,Camici:1997ij,Camici:1997sh,Ciafaloni:1998gs} tame the growth but other terms due to the parton recombination \cite{Gribov:1984tu} and multiple Pomeron exchanges  have to be taken into account in order to guarantee the unitarity of the scattering amplitudes. 

The Balitsky-Kovchegov equation  was derived in  \cite{Kovchegov:1999yj,Kovchegov:1999ua} in the dipole 
approach \cite{Mueller:1993rr} to high energy scattering in QCD and independently
in the formalism for the operator product expansion at high energy as an evolution of the Wilson line correlators with respect to the rapidity \cite{Balitsky:1995ub,Balitsky:1998kc,Balitsky:1998ya,Balitsky:2001re,Balitsky:2001mr}. This equation also emerges in the 
Color Glass Condensate formalism as a limit of the JIMWLK functional equation \cite{JalilianMarian:1997dw,JalilianMarian:1997gr,Iancu:2000hn,Iancu:2001ad,Ferreiro:2001qy},
see also \cite{Weigert:2000gi,Mueller:2001uk}. The BK equation was shown to be equivalent to the   BFKL evolution with   
the  triple Pomeron vertex  \cite{Bartels:1993ih,Bartels:1994jj} when the latter one is  restricted 
to the M\"obius space of functions \cite{Bartels:2004ef}.

Numerous analyses of BK equation which were performed up to date, see for example \cite{Lublinsky:2001bc,Lublinsky:2001yi,GolecBiernat:2001if,Armesto:2001fa,Braun:2001kh}, focused on finding the solutions to this equation under the assumption that the amplitude does not depend on the impact parameter but only on a dipole size and rapidity.   This assumption makes the equation relatively easy to solve (at least numerically). Usually one is justifying this approximation by considering  the scattering off a very large target, therefore bringing a large scale into the problem.  This results in  breaking the symmetry between the infrared and ultraviolet regions even in the case of the fixed coupling by generating a saturation scale which acts as a boundary and suppresses the diffusion into the infrared region \cite{GolecBiernat:2001if,Mueller:2002zm}. In the momentum space this approximation is equivalent to taking the forward limit in the evolution with the  simplifying approximation on the triple Pomeron vertex resulting in zero momentum transfer through the vertex in both reggeized gluon pairs  \cite{Bartels:2007dm}. On the other hand, the kernel of the equation  has the property that it is invariant with  respect to the translations, dilatations, rotations and inversions,  which are M\"obius transformations. Therefore the solutions to the nonlinear equation should also reflect this symmetry properties, at least in the leading-logarithmic,  fixed coupling limit. On a deeper level  it is related to the  M\"obius invariance of the $2\rightarrow 4$ reggeized gluon transition vertex  which was demonstrated explicitly \cite{Bartels:1995kf}.   
In the relation to the experiment, the detailed knowledge about the dynamics and expansion in impact parameter is of the utmost importance for the qualitative and quantitative description of the multiparticle production in hadronic  collisions.

In this paper we  analyze this equation numerically  including the full impact parameter 
dependence.  This is an extension of the previous work \cite{GolecBiernat:2003ym} where this type of analysis was performed, see also \cite{Gotsman:2004ra}.
 We significantly improve over \cite{GolecBiernat:2003ym} the numerical accuracy and technique which enables us to evolve the equation much faster and  to very high values of rapidity, of the order of $\sim 50$. In this way one can more accurately extract the asymptotic values of different exponents which govern the growth of the saturation scales in this equation. We confirm  the results of \cite{GolecBiernat:2003ym}, on the dependence of the scattering  amplitude as a function of dipole size and demonstrate that it vanishes for large dipole sizes. We also find the fast diffusion  of the solution in impact parameter space and recover the power tails.  The saturation scale both for small and large dipoles is extracted,  and the dependencies on the impact parameter and rapidity are found. The results of the solutions to the  equation in the leading logarithmic approximation (LL) are compared with the modified version of the equation proposed in \cite{Motyka:2009gi}. The modified version contains the cutoffs in rapidity which originate from kinematical constraints. These cutoffs contain kinematical constraints in only approximate way but we know  from the analysis of forward BFKL in momentum space that these constraints are known to reduce the speed of the evolution in a significant way \cite{Kwiecinski:1996td}, (for a  related analysis on
  impact parameter dependence in nonlinear equation and the energy conservation see \cite{Kormilitzin:2010at}. The BK without impact parameter dependence and with rapidity cutoffs was also analyzed in \cite{GolecBiernat:2001if,Chachamis:2004ab}). We also include running coupling in our analysis and find that the effect of the running coupling is quite different than in the case without the impact parameter. In this paper we consider a prescription for the running coupling with the external dipole as the scale as well as the prescription derived in 
 \cite{Balitsky:2008zza}.
The impact parameter dependent equation is extremely sensitive to the large dipole sizes and this is the region where the running coupling is very large and needs to be regularized by some other mechanism.

In this analysis we did not attempt to regularize the large dipole size region in any way. It is at present  totally unclear how confinement effects should be consistently included in the dipole formalism. Of course, for any phenomenological applications such cut should be included, perhaps similarly to what was done in \cite{Avsar:2006jy}. As we were interested in general properties of the evolution we did 
not attempt here to introduce additional cuts on large dipole sizes (via masses), which would interfere with the specific dynamics of the evolution.

The paper is organized in the following way. In the next section, Sec.~\ref{sec:bk} we 
briefly present the BK equation 
 and discuss the modified version which includes the cutoffs in rapidity. In Sec.~\ref{sec:numerics} we describe the numerical methods of finding 
the solution. In Sec.~\ref{sec:nob} we  first show the results of the solution without the impact parameter and extract the saturation scale for both the LL and the modified equation.   
In Sec.~\ref{sec:withb} we present the solutions with impact parameter. We discuss the form of the amplitude as a function of the dipole size, extract the saturation scales (both for small and large dipoles), and  discuss the form of the impact parameter profile 
which emerges in the evolution. We present the solutions both in the case of the LL and for the modified kernel.  Using the representation in terms of the conformal eigenfunctions we discuss the origins of different peaks in the amplitude as well as present estimates for the rapidity dependence of the small and large dipole saturation scales and the expansion radius in impact parameter. We also present the estimate of the cross section  
of the black disc radius and its dependence on the rapidity.  In Sec.~\ref{sec:runcoupling} we discuss the results with the running coupling, both  for the case without and with impact parameter dependence,
and for two different prescriptions of the running coupling.
Finally, in Sec.~\ref{sec:conclusions} we state our conclusions. 
 
\section{BK Kernel in LO and beyond}
\label{sec:bk}

In the leading logarithmic approximation in $\ln 1/x$, the nonlinear Balitsky-Kovchegov   
\cite{Kovchegov:1999yj,Kovchegov:1999ua,Balitsky:1995ub,Balitsky:1998kc,Balitsky:1998ya,Balitsky:2001re,Balitsky:2001mr}
evolution equation derived in dipole picture \cite{Mueller:1993rr}    has the following form  
\be    
\label{eq:BK}    
\frac{\partial N_{\xzb\xpb}}{\partial Y}\,=\,    
\overline{\alpha}_s\!    
\int\frac{d^2\xdb}{2\pi}    
\frac{(\xzb-\xpb)^2}{(\xzb-\xdb)^2(\xpb-\xdb)^2}\,    
\left[   
N_{\xzb\xdb}+N_{\xpb\xdb}-N_{\xzb\xpb}    
- N_{\xzb\xdb}\,N_{\xpb\xdb}   
\right]\, ,    
\ee    
where $\overline{\alpha}_s=\alpha_s N_c/\pi$ is the strong coupling constant.    
Here,  $N_{\xzb \xpb}\equiv N(\xzb,\xpb,Y)$ is the dipole-nucleus scattering amplitude, and $\xzb,\xpb$ are two-dimensional vectors of  the transverse position   
of the dipole  ends. Alternatively, one can introduce the vector    denoting the dipole size
$\rb_{01}=\xzb-\xpb$, and  the impact parameter $\bb_{01}=(\xzb+\xpb)/2$.    
Thus in general,    
the amplitude depends on the four degrees of freedom in transverse space and rapidity, $Y=\ln(1/x)$,    
playing the role of the  evolution parameter. The transverse part of the  LL kernel 
$$
\frac{dz}{z} \frac{d^2\xdb}{2\pi}    
\frac{x_{01}^2}{x_{02}^2 x_{12}^2} \; ,
$$
is conformally (M\"obius) invariant in 2-dimensions. Here, we introduced a more compact notation denoting
$\xb_{ij}\equiv \xb_i-\xb_j$, $x_{ij}=|\xb_{ij}|$ and $z$ is the longitudinal momentum fraction so that rapidity is $y=\ln 1/z$.
    
 To obtain the solution of this equation,   
one has to specify  an initial condition at $Y=Y_0$: $N_{\xzb\xpb}^{(0)}=N^{(0)}(\xzb,\xpb;Y=Y_0)$. 
   
    The amplitude $N_{\xzb \xpb}$ in (\ref{eq:BK}) is given by the following correlator    
\be   
\label{eq:Corr}   
N_{\xzb\xpb}\, = \,\frac{1}{N_c}\mbox{\rm Tr} \left<1-U^{\dagger}(\xzb) U(\xpb)\right> \, ,   
\ee   
where the trace is done in the color space,  and   
the eikonal factor $U$ is defined as the path ordered exponential with the $SU(N)$   
gauge fields (in the gauge $A_a^{-}=0$)   
\be   
\label{eq:eikonal}   
U(\xb)\, = \,  P\,  \exp\left\{i\int dx^{-}\,  T^a A_a^+(x^{-},\xb)\right\}\,.   
\ee   
The averaging $\left<...\right>$ in (\ref{eq:Corr}) is performed over an ensemble of classical gauge fields.   

The next-to-leading logarithmic corrections to the BK equation have been computed 
in \cite{Balitsky:2008zza,Balitsky:2009yp}.  
It would be instructive to perform the analysis of the BK equation with full NLL approximation. Due to the complicated form of this kernel  we are not trying to solve it here. What we consider instead, is the equation with two important modifications. The first one is the running coupling correction, which we analyze in Sec.~\ref{sec:runcoupling} of this paper. The second one is the subleading correction coming from the kinematical corrections, included in the form of the modified kernel proposed in \cite{Motyka:2009gi}.

The soft gluon approximation performed in \cite{Mueller:1993rr} was crucial in order to factorize the 
the branching kernel for the gluon emissions. In \cite{Motyka:2009gi}  a modified version of the kernel was proposed, which includes part of the subleading corrections coming from the kinematics. By taking into account an improvement in kernel  in which the value of energy denominator is obtained from the invariant mass of the produced gluon pair.
This leads to a new color dipole kernel which has the form

\be
\frac{dz}{z}\, z \;
  \frac{d^2 \xdb}{x_{01}^2}\, \left[K_1^2\left(\,{x_{02} \over x_{01}} \, \sqrt{z}\right)+K_1^2\left({x_{12} \over x_{01}} \, \sqrt{z}\right)-2\,K_1\left(\,{x_{02} \over x_{01}} \, \sqrt{z}\right)K_1\left({x_{12} \over x_{01}} \, \sqrt{z}\right)\frac{\xb_{02}\cdot\xb_{12}}{x_{02}x_{12}}\right] \; \;  ,
\label{eq:modkernel_0}
\ee
 where 
 $K_1$ is the modified Bessel function. The modified kernel becomes equivalent to the leading logarithmic kernel for $zx^2_{02} \ll x^2_{01}$ and $zx^2_{21} \ll x^2_{01}$ but differs significantly otherwise. To be precise, the production of larger dipole sizes is exponentially suppressed above the cutoff size which depends on the longitudinal momentum fraction of the soft gluon. 
The modified kernel introduces corrections at all orders beyond the leading logarithmic approximation.
The suppression of the large dipole emission implies also a suppression of the diffusion in the impact parameter. 
An alert reader might notice that the above modified kernel is not conformally invariant. This is due to the fact that in \cite{Motyka:2009gi} only part of the kinematical corrections were taken 
into account resulting in kernel (\ref{eq:modkernel_0}). In general there should be also a cutoff on small dipoles, similarly to what was done in  \cite{Avsar:2005iz}, which would yield conformally invariant kernel.  We postpone the derivation  and numerical analysis of such kernel to a later work.

In what follows we will analyze both forms of the equations numerically, taking into account also running coupling corrections. We shall demonstrate that the sensitivity to the subleading corrections strongly depends on the type of approximation one is working on, i.e. whether the solutions are impact parameter dependent or not.

\section{Numerical Methods}
\label{sec:numerics}
The BK equation was solved numerically by discretizing the scattering amplitude in terms of  variables  $(\log_{10}r,\log_{10}b,\cos \theta)$, where $\theta$ is the angle between impact parameter $\bb$ and dipole size $\rb$.  There is an additional angle, which reflects the overall orientation, we assumed that the solution is rotationally invariant. The amplitude $N(r,b,\cos\theta)$ was placed on a grid with dimensions $200_r\times200_b\times20_\theta$.  It was found that the number of points in the grid was less important than the absolute limits on the grid and the number of integration regions.  The limits of the grid were $10^{-8} \rightarrow 10^8$ in $r$ and $b$. We note that, the grid limits of $r$ and $b$ cannot not be set independently due to the correlations in dipole size and impact parameter.  The initial condition $N^{(0)}$ at rapidity $Y=0$ was chosen as in Ref.~\cite{GolecBiernat:2003ym} to be of the Glauber-Mueller \cite{Mueller:1999wm,Mueller:1999yb} form 

\begin{equation}
N^{(0)} = 1-e^{-10 r^2 e^{-\frac{b^2}{2}}} \; .
\label{eq:initial}
\end{equation}

This is the form of the scattering amplitude at the initial rapidity.  We note that in this analysis the constants are arbitrary and have not been fit to any experimental data.  The initial condition (\ref{eq:initial}) has the property that it reaches unity when the dipole size is large and it goes to zero for large values of  impact parameter by the steeply falling profile in $b$.  The amplitude is evolved in small steps $\Delta Y = 0.2$ as it was found that the smaller spacings had no significant effect on the accuracy of the evolution of the amplitude.  Below,  we denote integration over $\xdb$ simply by $\int_{\xdb}$ and this is understood to absorb the kernel as well as a measure $d^2 \xdb$ for ease of notation in this section. The basis of the method for the solution is the same as utilized in Ref.~\cite{GolecBiernat:2003ym} and we briefly outline it below. After one step of evolution from $Y_0$ to $Y = Y_0 + \Delta Y$   the equation (\ref{eq:BK}) becomes

\begin{equation}
N_{x_0 x_1}(Y_0+\Delta Y)=N_{x_0 x_1}(Y_0)+\int_{\xdb} \int_{Y_0}^{Y_0+\Delta Y} [N_{x_0 x_2}+N_{x_1 x_2}-N_{x_0 x_1}-N_{x_0 x_2}N_{x_1 x_2}] \; .
\label{eq:BFKLNUM}
\end{equation}

The lowest approximation for this equation is

\begin{equation}
N^{(1)}_{x_0x_1}=N^{(0)}_{x_0 x_1} + \Delta Y\int_{\xdb}[N^{(0)}_{x_0x_2}+N^{(0)}_{x_1x_2}-N_{x_0x_1}^{(0)}-N_{x_0x_2}^{(0)}N_{x_1x_2}^{(0)}] \; .
\end{equation}

The superscript 1 denotes the first approximation while the superscript 0 denotes the initial condition.  The accuracy of the equation is measured by the relative difference of the approximation defined by $|(N_{x_ix_j}^{(1)} - N_{x_ix_j}^{(0)})/N^{(0)}_{x_ix_j}|$. If this is less than some required accuracy  then the value of $N_{x_ix_j}^{(1)}$ is taken as the scattering amplitude at $Y=Y_0 + \Delta Y$. In our case  the relative accuracy was set to be equal $.02$.  If the accuracy condition is not satisfied, then we can find a second approximation by utilizing linear interpolation $N^{(2)}_{x_ix_j} = \frac{Y(N^{(1)}_{x_ix_j}-N^{(0)}_{x_ix_j})}{\Delta Y} + N^{(0)}_{x_ix_j}$.  By using this linear interpolation in the right hand side of \ref{eq:BFKLNUM} the expression for the next iteration can be attained

\begin{eqnarray}
N_{x_0x_1}^{(2)}&=&N^{(0)}_{x_0x_1} + \frac{\Delta Y}{2}\int_{\xdb}[N^{(0)}_{x_0x_2}+N^{(0)}_{x_1x_2}-N^{(0)}_{x_0x_1}-N^{(0)}_{x_0x_2}N^{(0)}_{x_1x_2}]\nonumber
\\&+&\frac{\Delta Y}{2}\int_{\xdb}[N^{(1)}_{x_0x_2}+N^{(1)}_{x_1x_2}-N^{(1)}_{x_0x_1}-N^{(1)}_{x_0x_2}N^{(1)}_{x_1x_2}]\nonumber
\\&+&\frac{\Delta Y}{6}\int_{\xdb}[N^{(1)}_{x_0x_2}-N^{(0)}_{x_0x_2}][N^{(1)}_{x_1x_2}-N^{(0)}_{x_1x_2}] \; .
\end{eqnarray}

If the accuracy condition mentioned earlier is still not satisfied this can be repeated by utilizing $N^{(n)} = \frac{Y(N^{(n-1)}-N^{(0)})}{\Delta Y} + N^{(0)}$.  The number of iterations required to reach this level of accuracy varies on how close one is to the initial condition, and the exact form of the equation (kernel used, running coupling form, etc).  Near the initial conditions it could take upwards of ten iterations but far away from the initial condition two or three iterations are usually sufficient to obtain the  accuracy assumed.  With this in mind an algorithm was written so that the iteration number was not fixed but only continued until the desired accuracy was reached.  Once the amplitude has been found for $Y = Y_0 + \Delta Y$ that takes the place of $N^{(0)}$ in the numerics and the procedure is repeated to find the amplitude at $Y = Y_0 + 2 \Delta Y$.  In this way the amplitude is evolved numerically to higher rapidities.

As was mentioned earlier the number of points in the grid was not as important as the number of integration regions the program executes to get a good solution.  For our limits it was found, that breaking the range into at least 20 integrations was needed.  This meant there would be an incredible number of function evaluations  and in order  to reach any useful rapidity on a standard dualcore machine would take upwards of a month.  The program was parallelized using the MPI (Message Passing Interface) library so it could be used with computer clusters to run the program.  Each run took approximately three days to run on 32 linked $3 \; {\rm GHz}$ processors and generated a total of 5 gigabytes of data.

\section{Results without impact parameter dependence}
\label{sec:nob}

In this section we briefly present the results of the numerical solution without impact parameter dependence. This was of course done in many previous works \cite{Lublinsky:2001bc,Lublinsky:2001yi,GolecBiernat:2001if,Marquet:2005zf}, but we repeat this analysis below for several reasons. First of all, we wanted to have benchmark for the comparison of the solutions. Secondly, we wanted to perform the analysis of the modified BK equation with Bessel function kernel, a computation not performed earlier in the literature.

The BK equation can be evaluated without impact parameter or angular dependence  and this greatly reduces the time needed to evolve the scattering amplitude.  The initial condition used in this case is taken to be

\begin{equation}
N^{(0)} = 1 - e^{-r^2} \; , 
\label{eq:OneDInitial}
\end{equation}
without an impact parameter profile.
This form goes to unity as the dipole size gets large and has a narrow transition region from 0 to 1.  The absolute limits on the dipole size here are $10^{-10} \rightarrow 10^{10}$ and the number of points in the grid is 5000.  These results are for fixed coupling $\bar{\alpha}_s = 0.1$, the case with running coupling is considered in section Sec.~\ref{sec:runcoupling}.

\begin{figure}
\includegraphics[angle=270,width=3.1in]{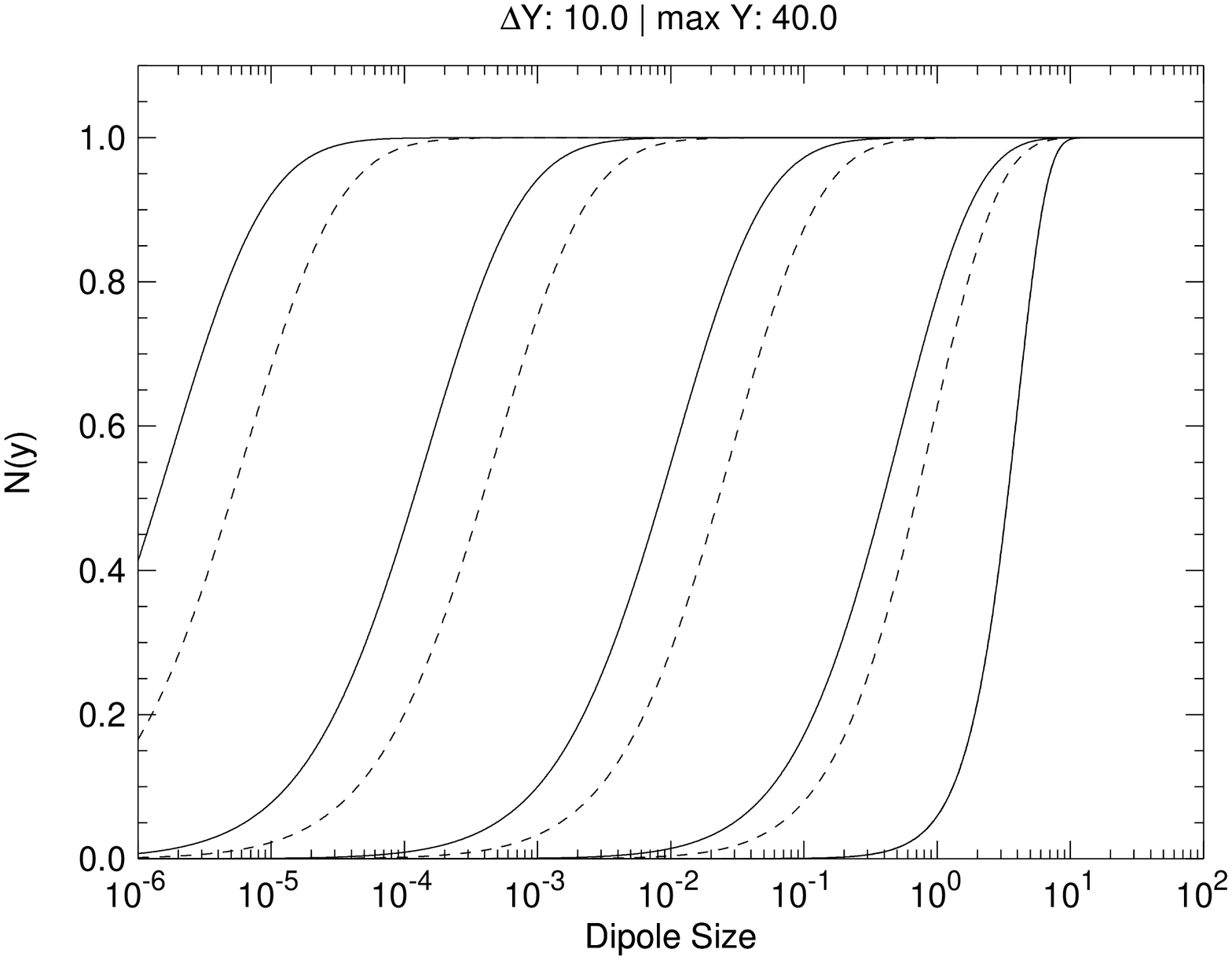}\hspace*{0.5in}
\includegraphics[angle=270,width=2.9in]{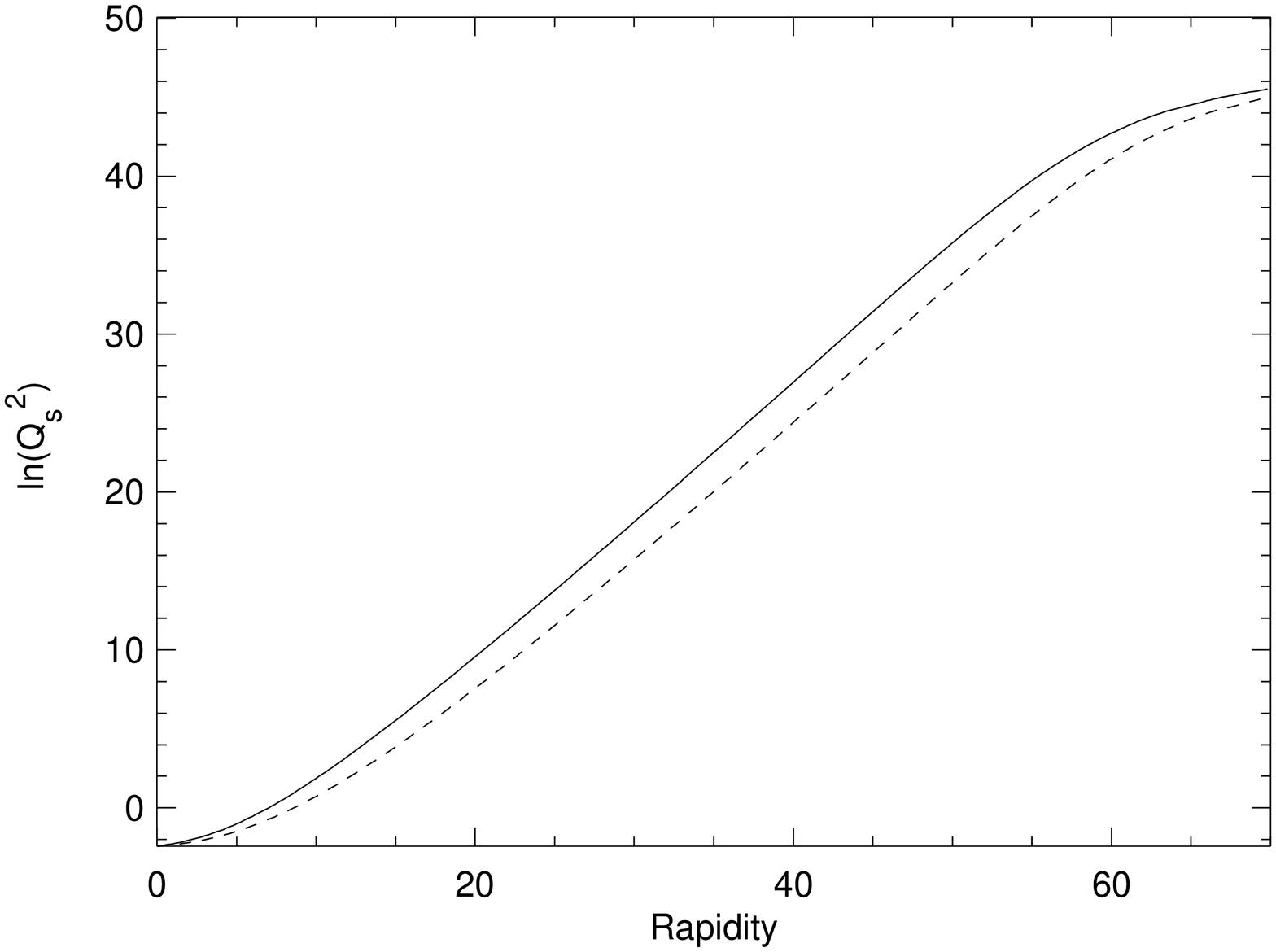}

\caption{Left: Comparison between the LO kernel (solid lines) and the kernel with Bessel functions, (\ref{eq:modkernel_0}) (dashed lines) at constant rapidities.  The solid line at the far right is the initial condition for both cases and each line to the left represents an evolution in ten units of rapidity to a maximum of $Y=40$. \\
Right: plot of the logarithm of the saturation scale as a function of rapidity.  The solid line represents the evaluation with LO kernel and the dashed line represents the solution with Bessel kernel.  The slope of the curve gives $\lambda_s$ and the three regimes of evolution can be seen. Calculation done for the fixed coupling case and without the impact parameter.
\label{fig:OneDBess}}
\end{figure}

The solid lines shown in left graph in Fig.~\ref{fig:OneDBess} correspond to a set of scattering amplitudes found from the LO BK equation at constant rapidity.  The result has the well known form of a traveling wave with rapidity playing the role of the time in the evolution. 
As rapidity increases the front moves towards the small values of dipoles.

Next, we performed the evolution of the modified equation with the Bessel function kernel and the solutions are illustrated by the dashed lines in Fig.~\ref{fig:OneDBess}.
The normalization of the amplitude in the dilute regime is much smaller than the LO case.
  This is to be expected as the $K_1$ functions fall off faster than the power-like LO kernel.  The speed of this evolution can be quantified by the evaluation of the  saturation scale.  The equation to define saturation scale is

\begin{equation}
\langle N(r=1/Q_s,Y)\rangle \;  = \;  \kappa \; , 
\end{equation}

where $\kappa$ is a constant between 0 and 1. It determines   the point at which nonlinearities begin to become important in the evolution equation.  In all the analysis in this paper $\kappa = 0.5$ was chosen.  Based on results of \cite{Munier:2003sj} we expect the saturation scale to have the following form

\begin{equation}
Q_s^2(Y) \; = \; Q_0^2 \,\exp{\left[\bar{\alpha}_s\left( \frac{\chi(\gamma_c)}{1-\gamma_c}Y - \frac{3}{2(1-\gamma_c)} \ln Y \right)\right]} \; ,
\label{eq:muniersat}
\end{equation}
where $\chi(\gamma)$ is the BFKL kernel eigenvalue, and $\gamma_c=0.62$. This gives $\frac{\chi(\gamma_c)}{1-\gamma_c}=4.88$ and $\frac{3}{2(1-\gamma_c)}=2.39$ for the LO kernel.
The second term in the exponent is a correction of less than 10 percent compared to the first term for the rapidities which we are considering  so we take the saturation scale to be parameterized in this paper by

\begin{equation}
Q_s^2 = Q_0^2 e^{\lambda_s \bar{\alpha}_s Y} \; ,
\label{eq:satparam}
\end{equation}

where $\lambda_s$ is extracted from the numerical solution and $Q_0$ is a normalization term.  The extracted value for the saturation exponent is $\lambda_s=4.4$ for both the LO and Bessel function kernels. The exponent was extracted both for $\asb=0.1$ and $\asb=0.2$
and it was found to be the same.
The values of $\lambda_s$ are evaluated in the region of rapidities where the exponent is constant, i.e. nearly asymptotic.
 We observed that for the lower rapidities,  near the initial condition the exponent slowly increases and at very large rapidities the exponent begins to decrease.  The latter effect is  due to the absolute limits of the dipole size (or the box in which the equation is being solved). This strongly  affects the solution, and one of the effects is to slow down the evolution.  This effect can be clearly seen in Fig.~\ref{fig:OneDBess} where the three regimes are distinct.  The first regime is where  there are a lot of preasymptotic effects and there is strong dependence on the initial condition and the speed of evolution increases (approximately between 0 and 34 units of rapidity), the  regime where evolution is a constant (approximately between 34 and 47 units of rapidity) and finally the regime where box size limits growth (rapidity ~47+).  These values clearly shift with box size and they also change depending on the value of $\bar{\alpha}_s$ and to a lesser extent which kernel is used.  
 
The extracted value of the exponent of the saturation scale is consistent with the theoretical predictions of \cite{Munier:2003sj}. What is however surprising at first, is the fact that the asymptotic value of $\lambda_s$ is nearly the same for the modified version of the equation.
This is also evident by  investigating  right plot in Fig.~\ref{fig:OneDBess}
where the two saturation scales grow with rapidity in almost exactly the same way.
 This effect  can be easily understood  by inspecting   the form of the modified kernel

\begin{equation}
Q^2_{01}\left[K_1^2(Q_{01}x_{02})+K_1^2(Q_{01}x_{12})-2K_1(Q_{01}x_{02})K_1(Q_{01}x_{12})\frac{x_{02}\cdot x_{12}}{|x_{02}||x_{12}|}\right] \; .
\label{eq:modbk}
\end{equation}

Here $Q_{01} = \sqrt{z}/x_{01}$ and $z = e^{-y}$.  When $Q_{01}$ is small and the arguments of the Bessel functions are small this expression reduces to the LO kernel.  At large rapidities where the exponent $\lambda_s$ is evaluated,  $z$ is small and thus $Q_{01}$ is also small.
The difference between the modified kernel and the LO kernel comes from the regime   where the dipole sizes $x_{12}, x_{02}$ are large, that is when
$$
x_{12}, x_{02} \gg \frac{1}{Q_{01}} \;.
$$
In this large dipole regime (and for large rapidities),  the scattering amplitude is close to unity and  the right hand side of the BK equation goes to zero so there is no contribution from that regime. 
This means that the effect of the infrared modification of the kernel of the type (\ref{eq:modbk}) is negligible when the nonlinear evolution is considered. The modified kernel (\ref{eq:modbk})
was shown to reproduce the double-logarithmic terms in the exact NLL calculation. 
These type of  terms are known to come from the scale choice in the amplitude, 
\cite{Salam:1998tj,Ciafaloni:2003rd,Vera:2005jt}.

Two comments are in order here. The first observation is that, as we shall see later, the presence of the  impact parameter changes the dynamics significantly, and in this case the dependence on the form  of the kernel is more pronounced. In particular, in this case we do find the differences in the evolution speed due to the cutoffs in the infrared.  The second comment is that, in order to address the question of the scale choices one needs to take into account the corrections beyond the BK equation. BK equation is highly asymmetric with respect to the target and projectile and therefore cannot address the problems of all the scale choices. The corrections would necessarily have to include the Pomeron loops in order to make the evolution symmetric with respect to the target and projectile.   Such formulation has been  implemented in a Monte Carlo approach  \cite{Avsar:2005iz,Avsar:2006gw,Avsar:2006jy,Avsar:2007xh} with Pomeron loops effectively taken via so-called dipole swing and with cutoffs both in the infrared and ultraviolet regions.
\section{Results with impact parameter dependence}
\label{sec:withb}
\subsection{Dependence on the dipole size}

\begin{figure}
\centering
\subfigure{\label{fig:Rdepa}\includegraphics[angle=270,width=0.48\textwidth]{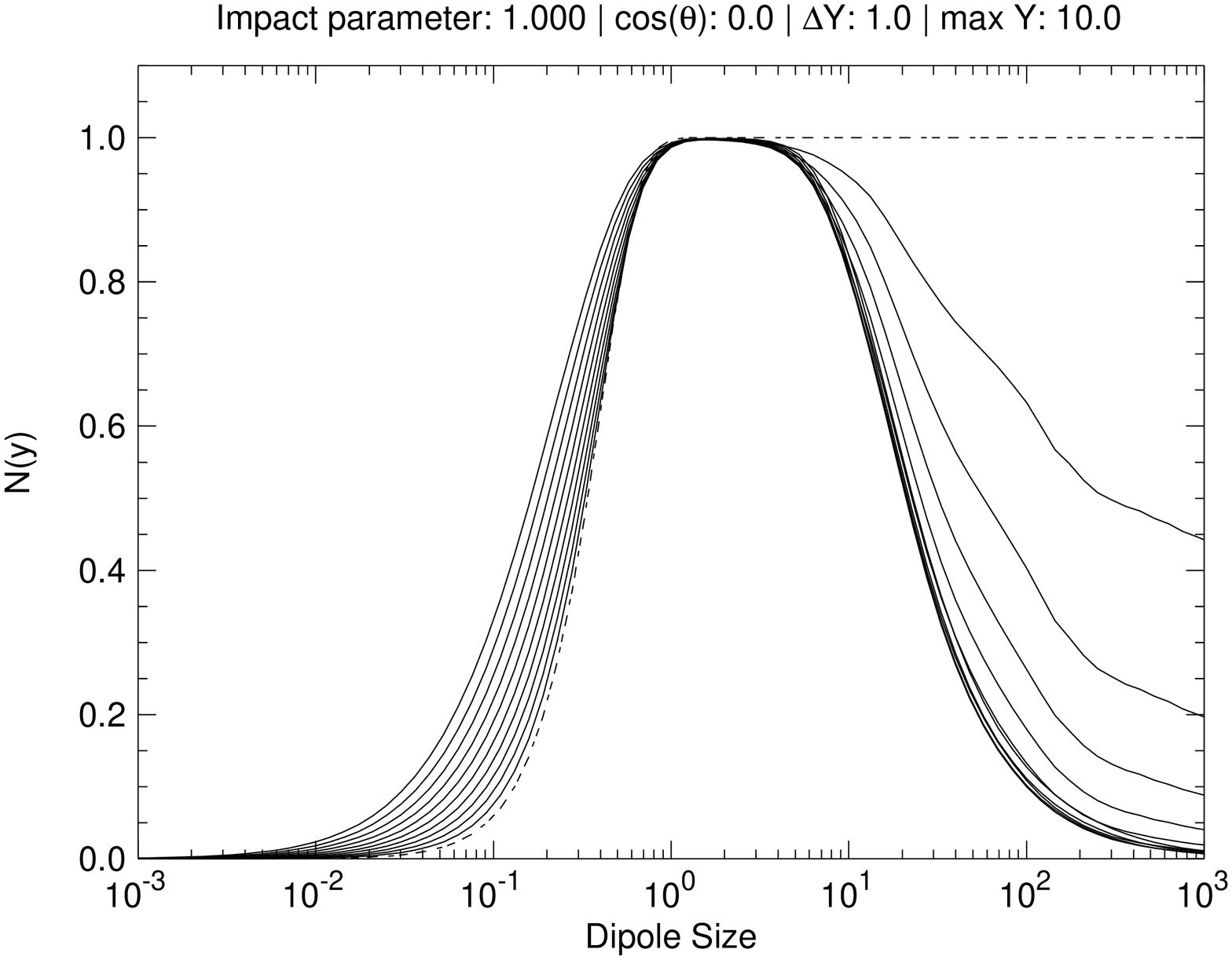}}
\subfigure{\label{fig:Rdepb}\includegraphics[angle=270,width=0.48\textwidth]{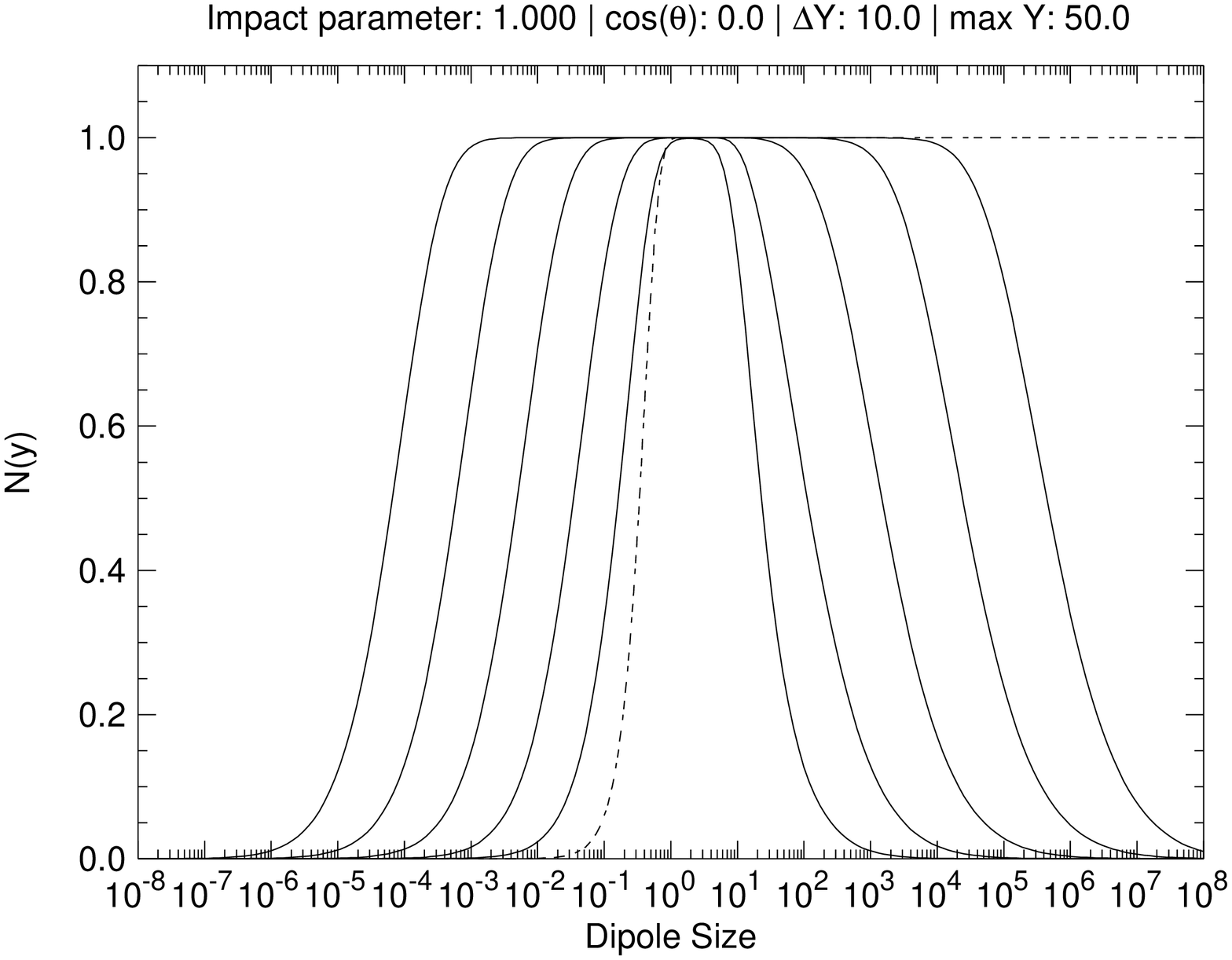}}
\caption{Graphs of the scattering amplitude $N(x,b,\theta,Y)$ at various constant rapidities for the calculation with impact parameter dependence. Left: the individual lines correspond to the rapidity intervals $\Delta Y=1.0$ up to rapidity of $Y=10$. Right: the individual lines correspond to the rapidity intervals of $\Delta Y=10$ and the maximal rapidity is $Y=50$. The dotted line represents the initial condition which is the same in both graphs.  Impact parameter and angle are fixed for both graphs at $b = 1.0$ and $\cos(\theta) = 0$.  }
\label{fig:RdepUcor}
\end{figure}

In this section we discuss the results which include the impact parameter dependence as well as the angular dependence between the dipole sizes and the impact parameter orientation.
Let us  first investigate the dependence of the scattering amplitude on the dipole size with fixed impact parameter value. Fig.~\ref{fig:RdepUcor} shows that case of the LO simulation with $\asb=0.1$.
  The scattering amplitude at small dipole sizes evolves in a similar manner as in the case without the impact parameter. 
 
 At large dipole sizes the situation is drastically changed with respect to the translationally invariant case.
 Here, we observe that the amplitude drops down from the initial distribution and forms a second evolution front.
   This drop  is very rapid compared to the expansion evolution of the small dipole size regime or the large dipole size regime at higher rapidities.  The evolution of the front at large dipole sizes can be best seen in Fig.~\ref{fig:Rdepb} where the steps in rapidity are greater.  
   As discussed in \cite{GolecBiernat:2003ym} there is a clear physical reason for this effect.
   For a large dipole, its end points are in the region where there is no gauge field. In this situation the gauge field correlator
  
\be   
\label{eq:Corr2}   
N(\xb,\yb)\, = \,\frac{1}{N_c}\mbox{\rm Tr} \left<1-U^{\dagger}(\xb) U(\yb)\right> \, ,   
\ee   
vanishes because $U(\xb)\simeq U(\yb)\simeq 1$.
In this case the dipole is larger  than any of the other scales in the problem, including the impact parameter of the collision.
   In other words this situation corresponds to the setup of very large dipole scattering on a localized target, and in this case the dipole misses the target, resulting in the vanishing scattering amplitude. This effect is not present in the case where impact parameter dependence is neglected simply because this approximate case corresponds to the infinitely large target.

It is interesting to investigate the dipole size dependence of the amplitude at large values of the impact parameter. There the initial condition,
sets the amplitude to zero. The evolution quickly changes this value, and a unique feature of the solution develops. Namely, a peak is formed with the center at  the dipole size value which is exactly twice the impact parameter value. The peak grows until saturation is reached and the evolution of the fronts proceeds to the infrared and ultraviolet regions. The discussion about the origin of peaks is given in subsection ~\ref{sec:conformal}.
\begin{figure}
\centering
\subfigure[ $ \; \;b=0.11$]{\label{fig:Rdepa2}\includegraphics[angle=270,width=0.44\textwidth]{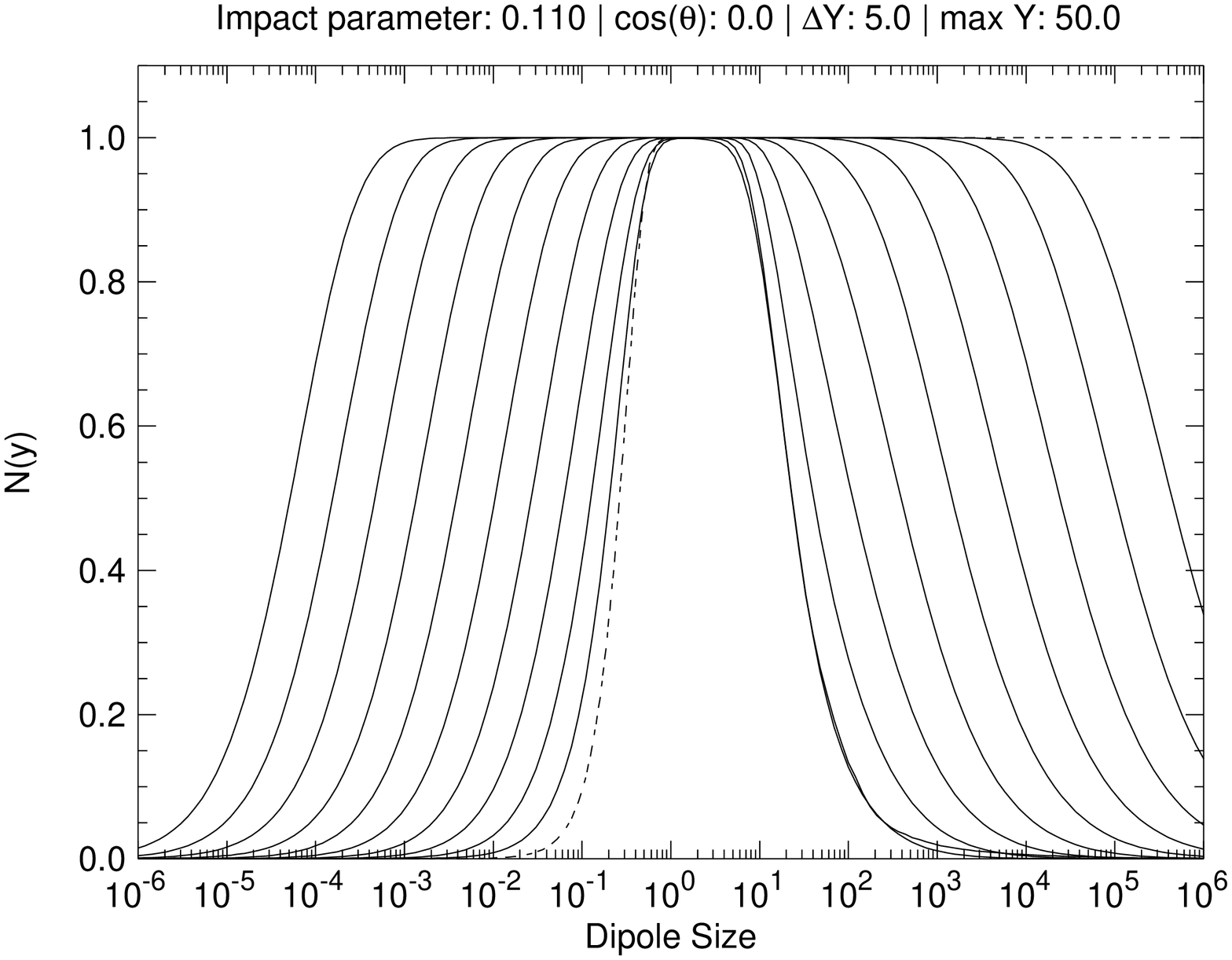}}
\subfigure[ $\; \; b=100$]{\label{fig:Rdepb2}\includegraphics[angle=270,width=0.45\textwidth]{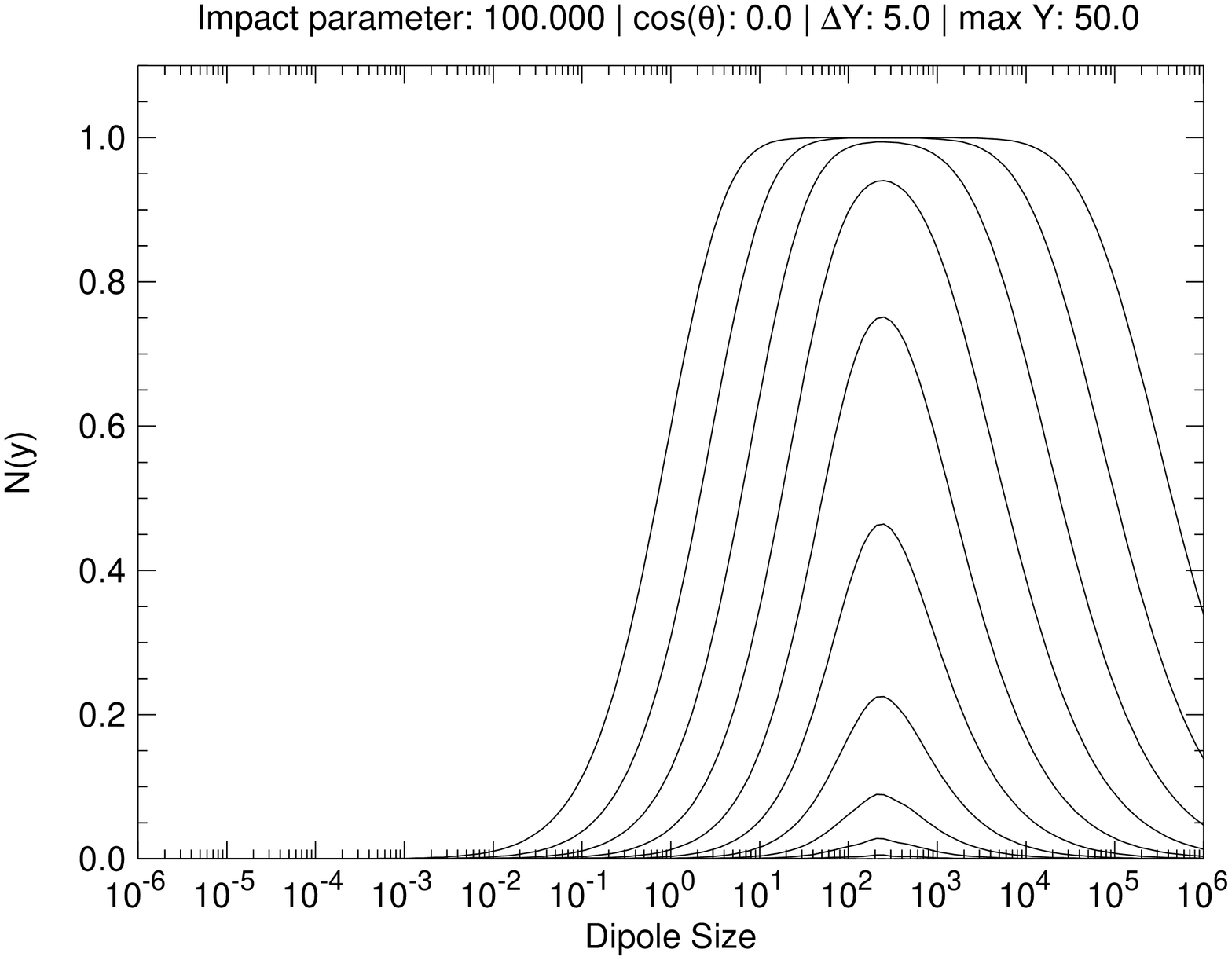}}
\caption{Graphs of the scattering amplitude at different constant rapidities for two different impact parameter values.  The dotted line represents the initial condition which is the same in both graphs, however the initial condition is near zero on the second graph with $cos(\theta) = 0$.  Each line past the initial condition represents five rapidity units to a maximum of 50.}
\label{fig:RdepUcor2}
\end{figure}

\begin{figure}
\centering
\subfigure{\label{fig:RdepCompa}\includegraphics[angle=270,width=0.48\textwidth]{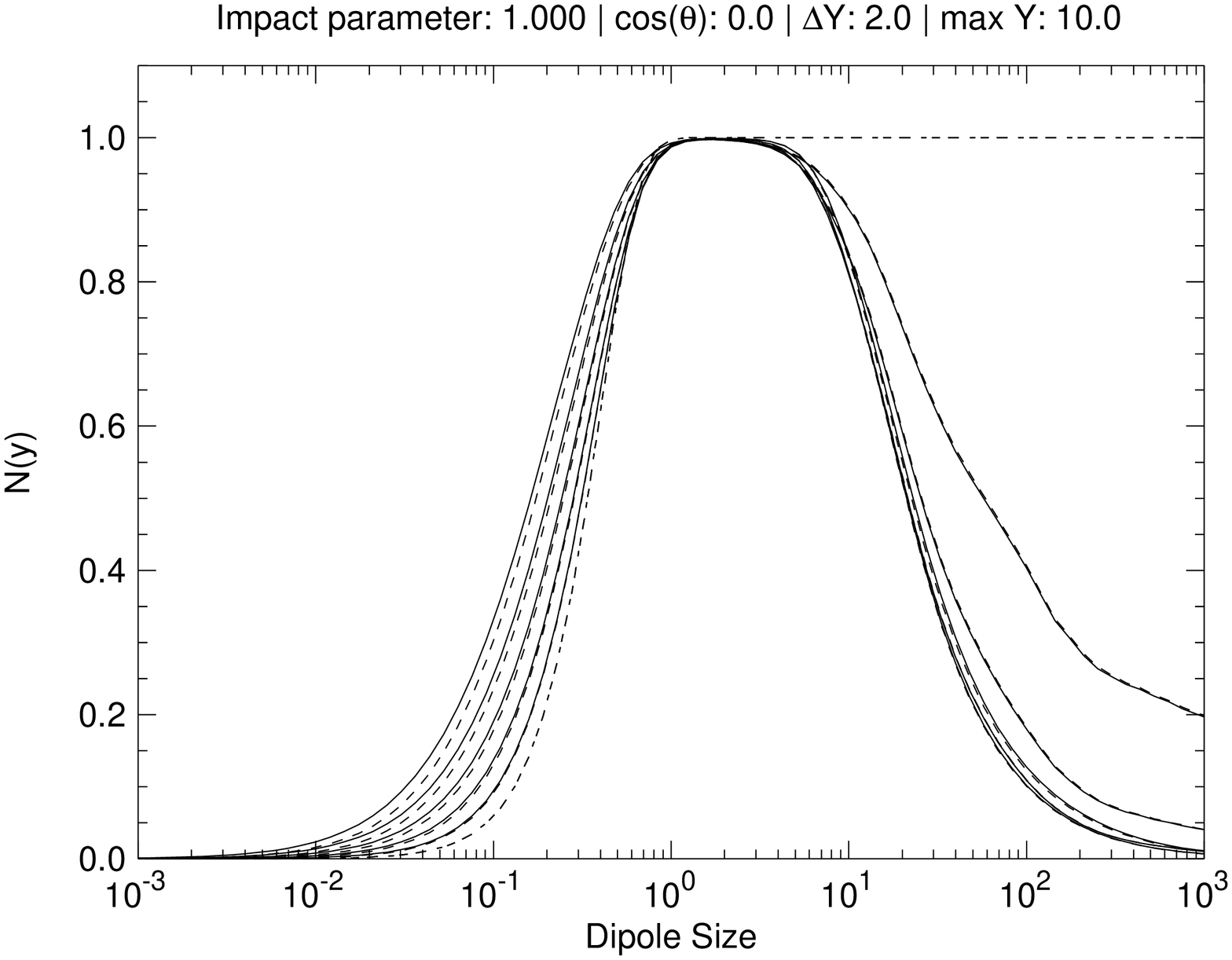}}
\subfigure{\label{fig:RdepCompb}\includegraphics[angle=270,width=0.48\textwidth]{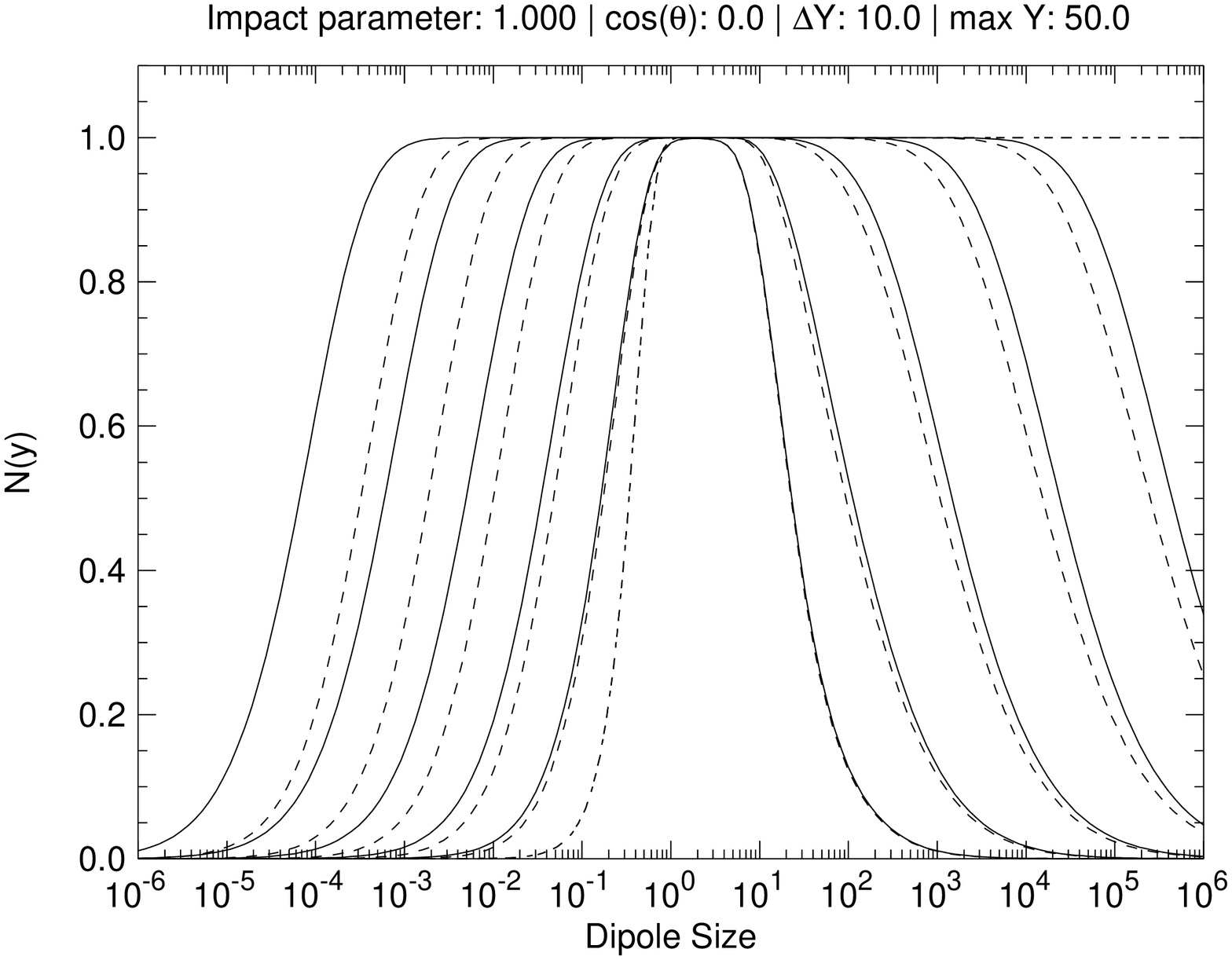}}
\caption{Graphs of the scattering amplitude as a function of the dipole size at various constant rapidities for fixed impact parameter $b=1.0$ and angle  $\cos(\theta) = 0$. Solid lines  are for  the LO  kernel and the dashed lines correspond to the Bessel kernel.  The initial distribution is equivalent for both kernels and is represented by the dotted-dashed line. On the left graph  each line represents a change in two units of rapidity to a maximum of ten and on the right graph each line represents a change in ten units of rapidity to a maximum of fifty.}
\label{fig:RdepComp}
\end{figure}

We have repeated this analysis for the case of the modified  kernel. 
As can be seen in Fig.~\ref{fig:RdepCompa} the evolution in the small dipole regime is now quite different from the LO kernel. The evolution is significantly slower in this regime for the Bessel function kernel.
Interestingly, the evolution in the large dipole region is slowed too but not as much as for small dipoles.
This is best illustrated in Fig.~\ref{fig:RdepCompb} where the evolution has been performed to large rapidities.
  The origins for this behavior are discussed in detail in the section on the saturation scale and evolution speed.

\subsection{Impact parameter profile of the scattering amplitude}

\begin{figure}
\includegraphics[angle=270,width=3.0in]{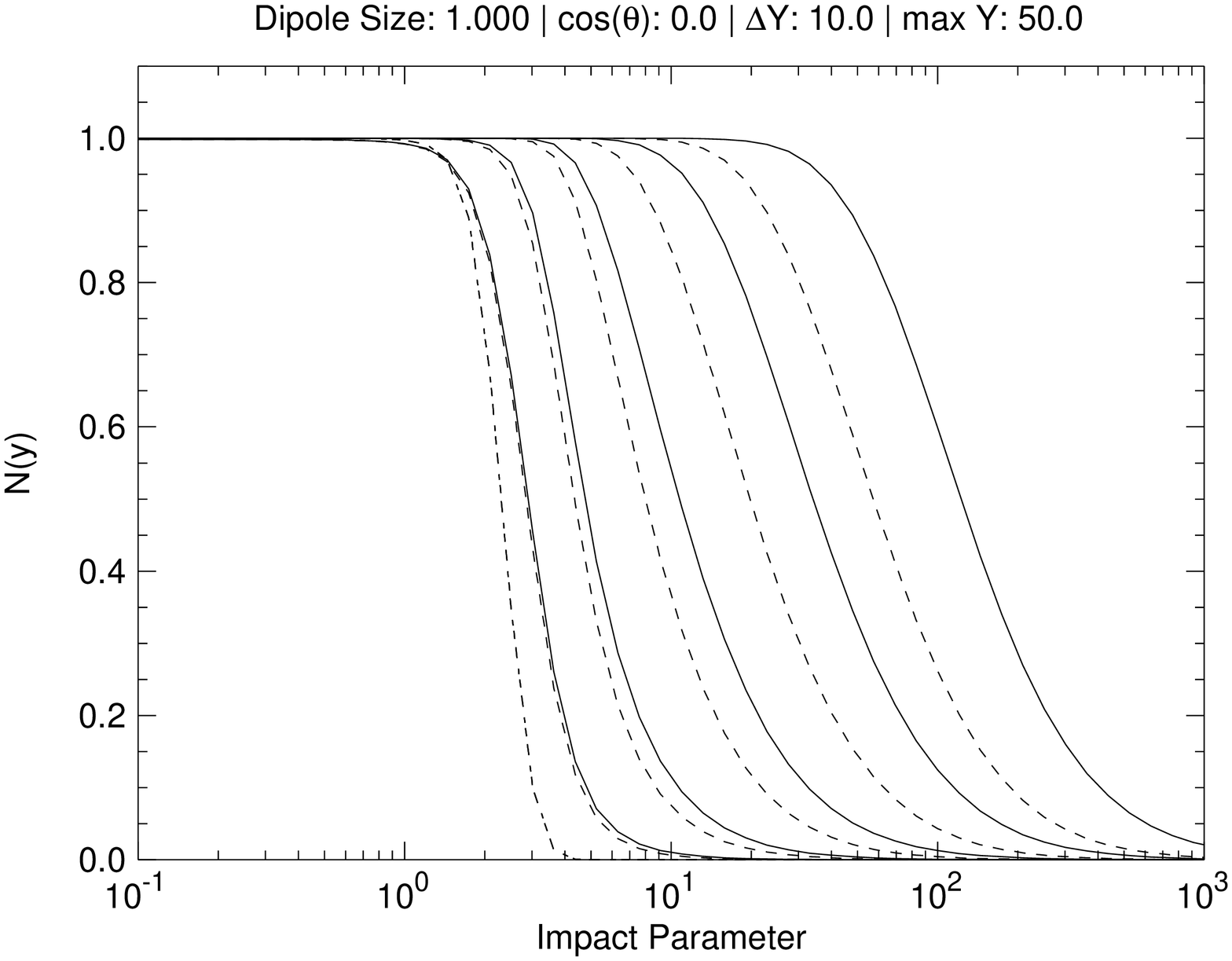}
\includegraphics[angle=270,width=3.0in]{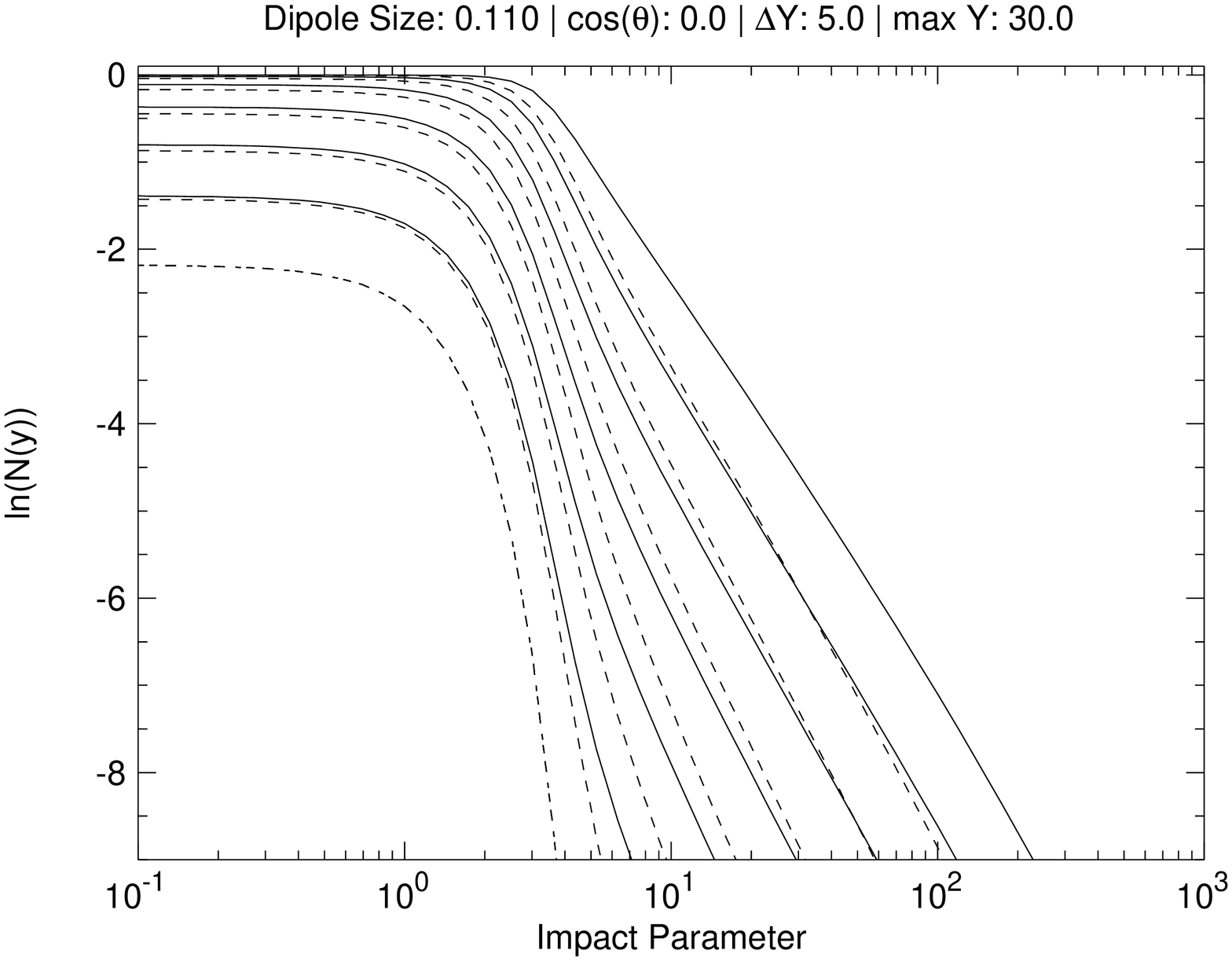}
\caption{Graph of scattering amplitude as a function of impact parameter for fixed dipole size $r=1.0$.  The solution with the case of the LO kernel is plotted as a solid line and with the modified kernel 
(\ref{eq:modbk}) as a dotted line.  The dotted-dashed line on the left is the initial condition.  Each line thereafter represents an increase in rapidity of ten units to a maximum of fifty. Right plot: the same but for the dipole size $r=0.11$ and in logarithmic scale for the amplitude.\label{fig:UcorCompB1}}
\end{figure}

Dependence of the dipole amplitude on the  impact parameter is illustrated  in Fig.~\ref{fig:UcorCompB1}.
The leftmost dashed-dotted line is the initial condition Eq. \ref{eq:initial} which has a very steep profile 
in impact parameter. The evolution of the scattering amplitude towards large values of impact parameter  follows the diffusion of large dipoles. The speed of this evolution can be extracted numerically and is determined by the expansion of the black disc radius. We will discuss this quantity in detail in the next section.

\begin{figure}
\centering
\subfigure[$\; \;\cos(\theta) = 1.0,-1.0$]{\label{fig:AngleCompR1}\includegraphics[angle=270,width=0.49\textwidth]{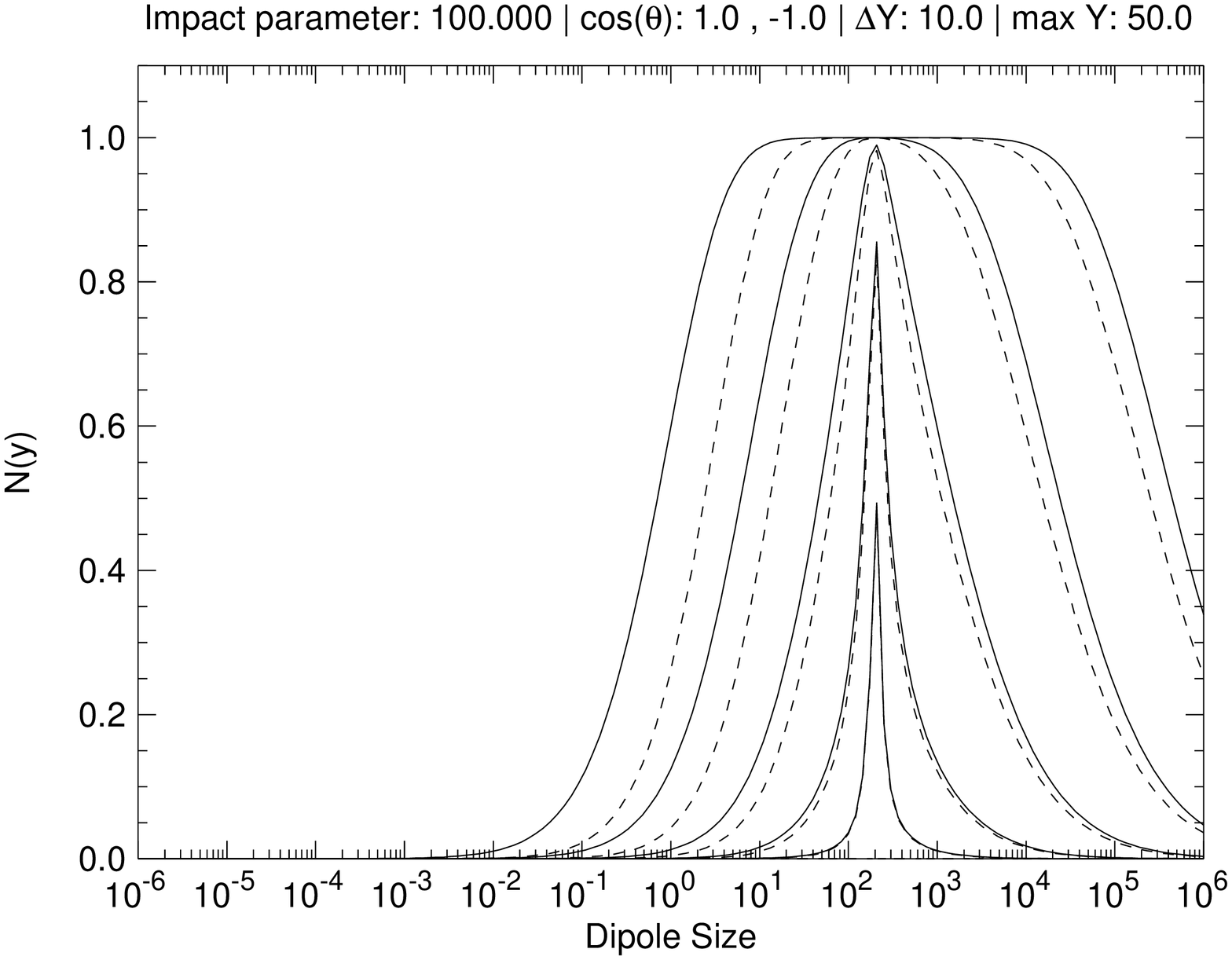}}
\subfigure[$\; \; \cos(\theta) = 0.0$]{\label{fig:AngleCompR2}\includegraphics[angle=270,width=0.45\textwidth]{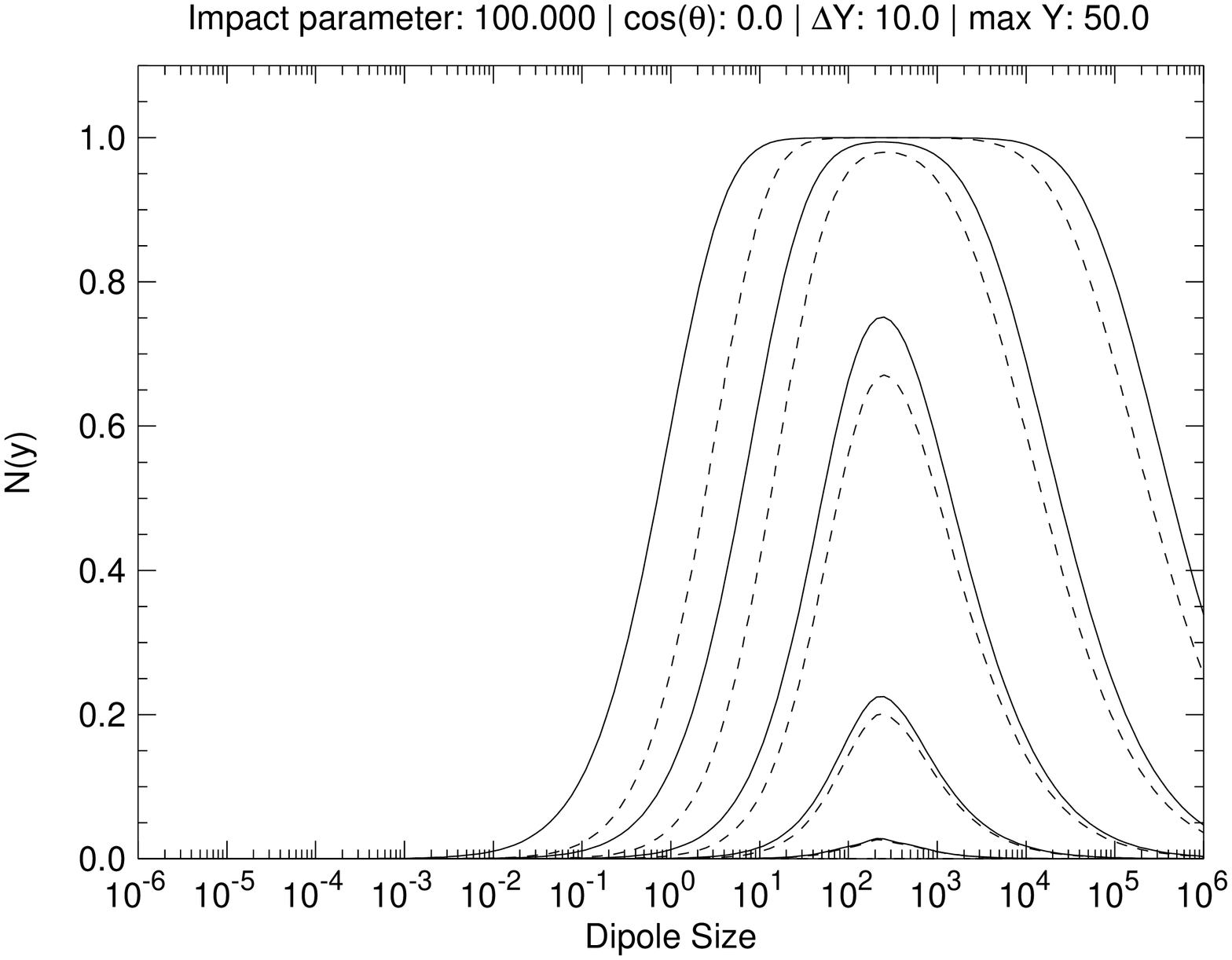}}
\caption{Graphs of scattering amplitude versus dipole size for fixed impact parameter $b=100.0$ and various rapidities and angles.  The initial condition is the same in all graphs and it is near zero, each curve represents an increase in ten units of rapidity to a maximum of fifty.  The LO kernel (solid lines) and the Bessel kernel (dotted lines) are plotted on the same graph.}
\label{fig:AngleCompR}
\end{figure}

\begin{figure}
\centering
\subfigure[$cos(\theta) = 1.0,-1.0$]{\label{fig:AngleCompB1}\includegraphics[angle=270,width=0.49\textwidth]{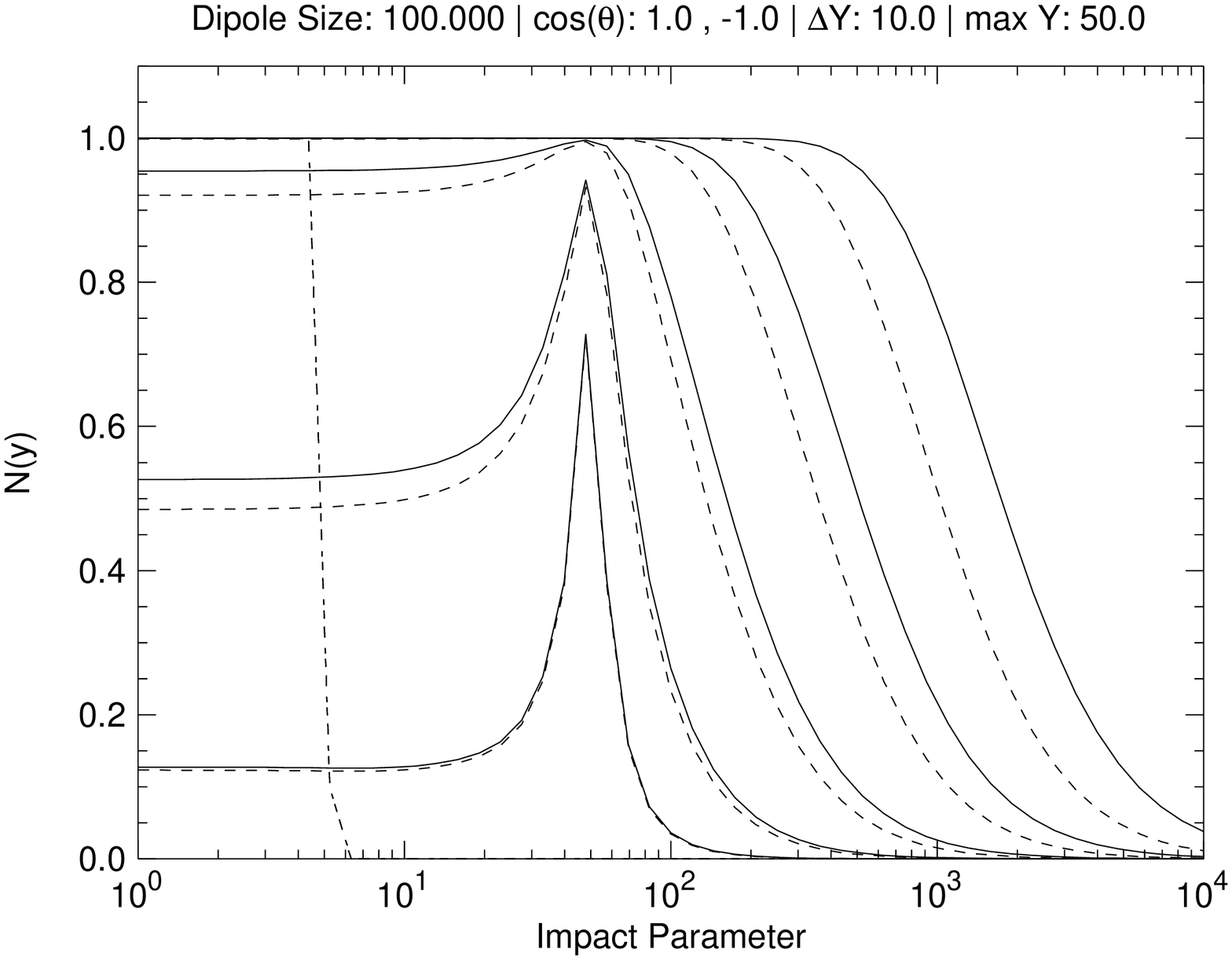}}
\subfigure[$cos(\theta) = 0.0$]{\label{fig:AngleCompB2}\includegraphics[angle=270,width=0.46\textwidth]{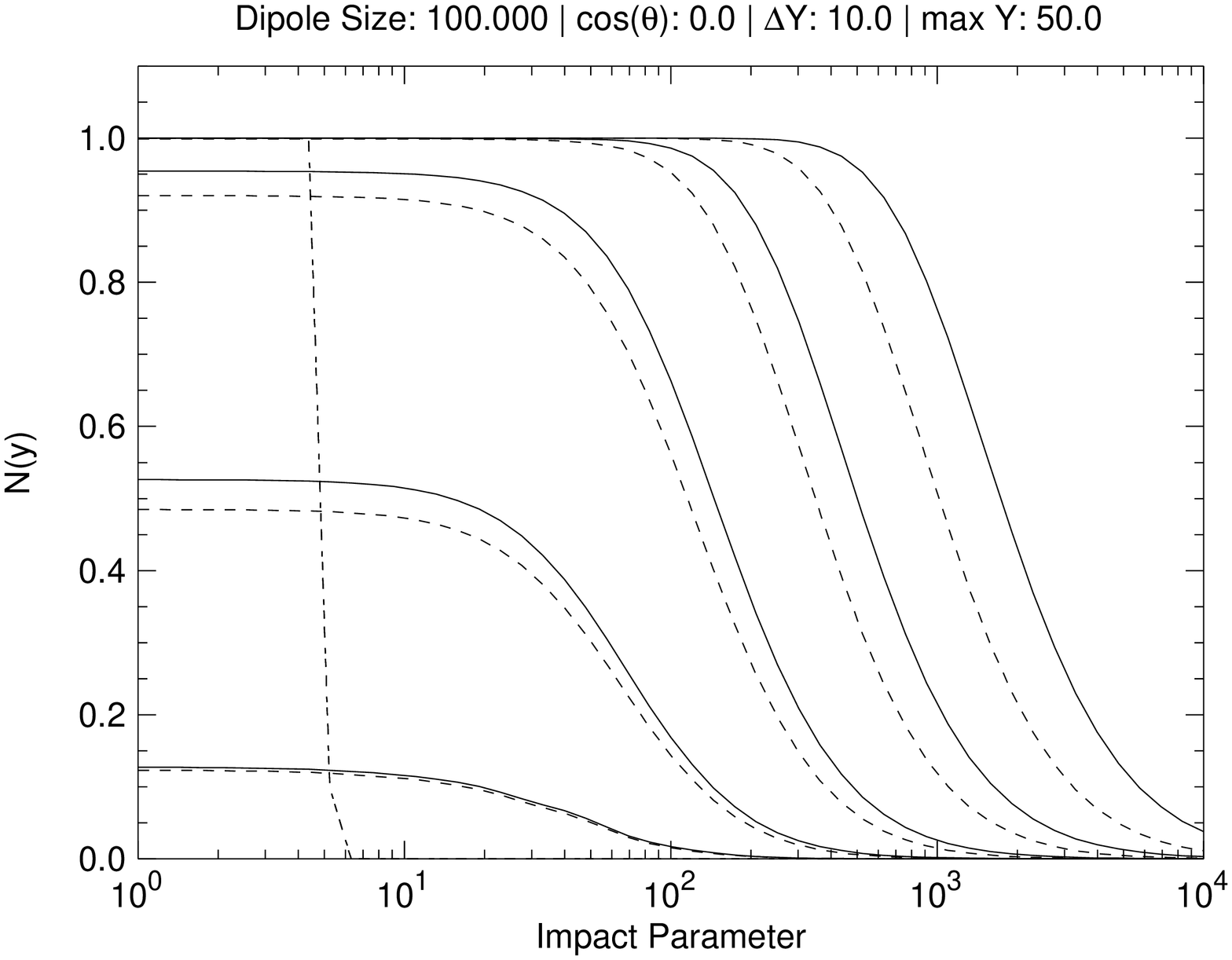}}
\caption{Graphs of scattering amplitude versus dipole impact parameter for constant dipole size $r=100.0$ and various rapidities.  The initial condition is the same in all graphs and it is the steeply falling dotted-dashed curve, which is the same for both the evolution with LO kernel (solid lines) and the Bessel kernel (dotted lines).  Each curve represents an increase in ten units of rapidity to a maximum of fifty.}
\label{fig:AngleCompB}
\end{figure}

\begin{figure}
\centering
\subfigure[$r = 1.0$ $b = 1.0$]{\label{fig:AnglePlots1}\includegraphics[angle=270,width=0.49\textwidth]{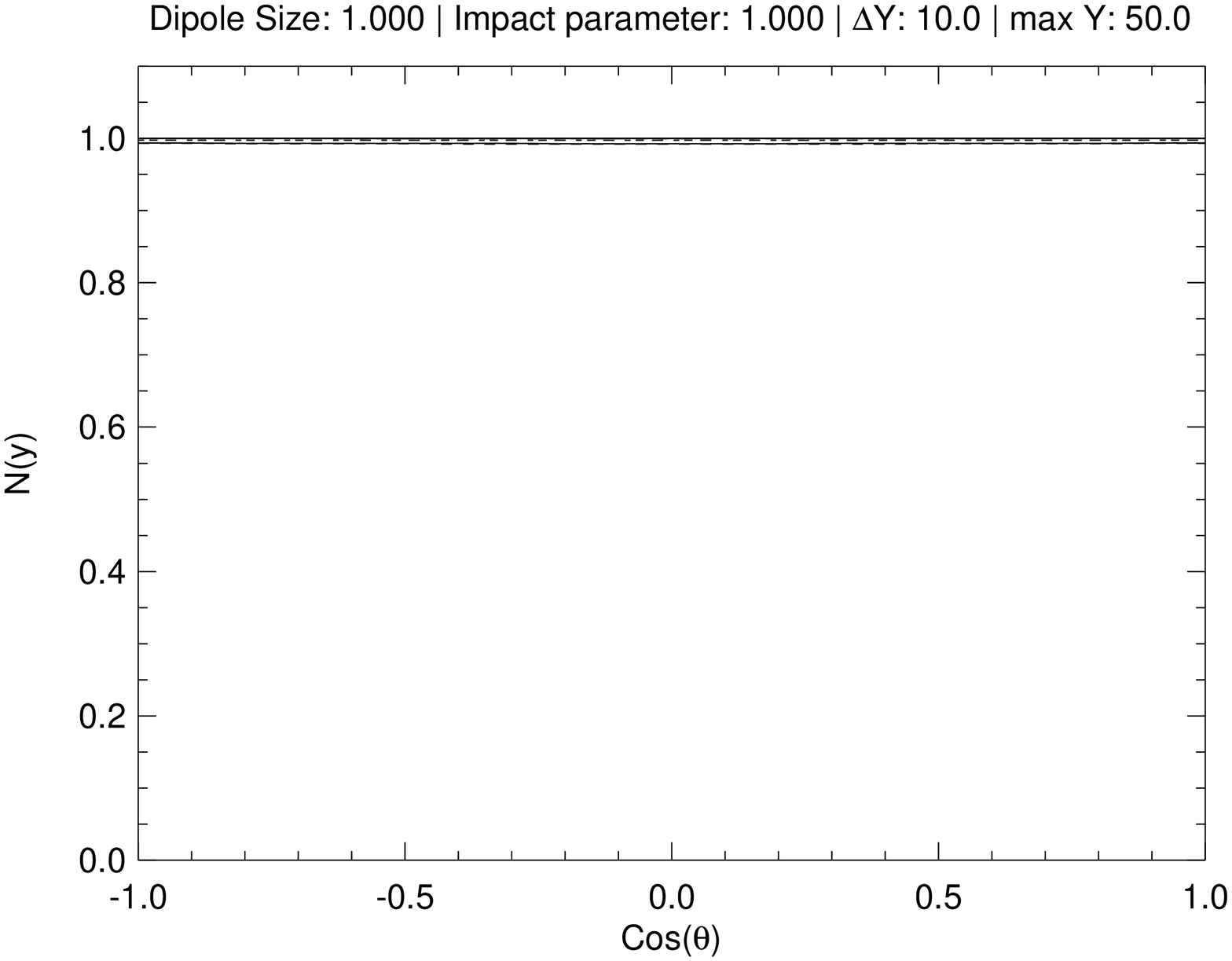}}
\subfigure[$r = 100.0$ $b = 47.8$]{\label{fig:AnglePlots2}\includegraphics[angle=270,width=0.49\textwidth]{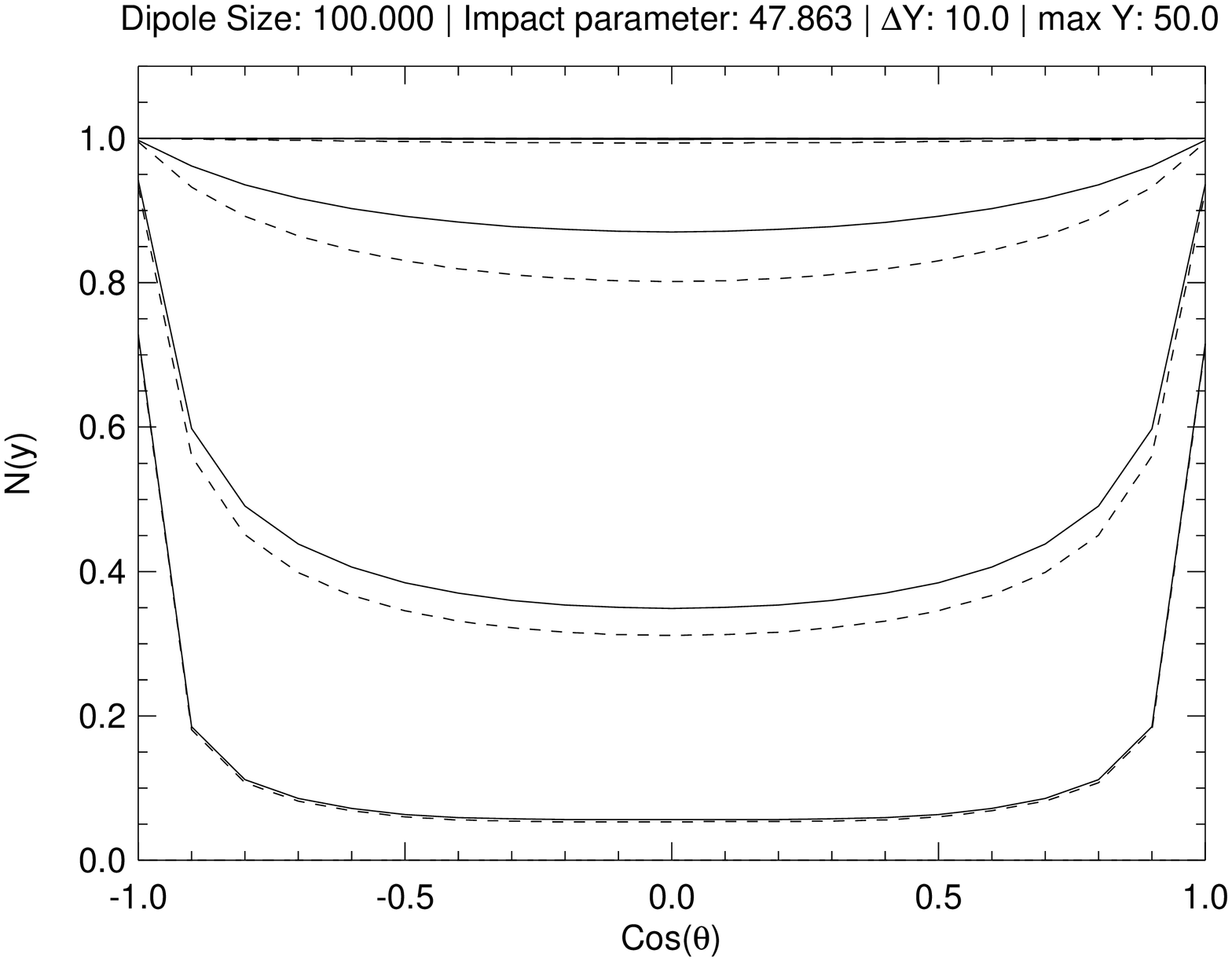}}
\caption{Graphs of scattering amplitude versus angle.  Both have the LO kernel (solid line) and the Bessel kernel (dashed line) graphed on them.  The first graph shows no angular dependence and the second shows a marked increase at values of $cos(\theta) = 1.0,-1.0$ which accounts for the difference in peaks in Figs. (\ref{fig:AngleCompB},\ref{fig:AngleCompR})}
\label{fig:AnglePlots}
\end{figure}

Evolution in impact parameter shows a marked change in profile from the steeply falling exponential in the initial condition.  This is better illustrated  in right plot in  Fig. \ref{fig:UcorCompB1} where we replot the impact parameter using the logarithmic scale in scattering amplitude.  The profile changes from the exponential to a power tail at small scattering amplitudes. This can be seen as an 'ankle' in the curves of constant rapidity.  The origin of this power-like tail was discussed in  detail in Ref.~\cite{GolecBiernat:2003ym}.  These power tails are also present in the modified kernel. In the latter case however there is a slower evolution of the profile towards the large values of impact parameters.
\begin{figure}
\includegraphics[angle=0,width=2.6in]{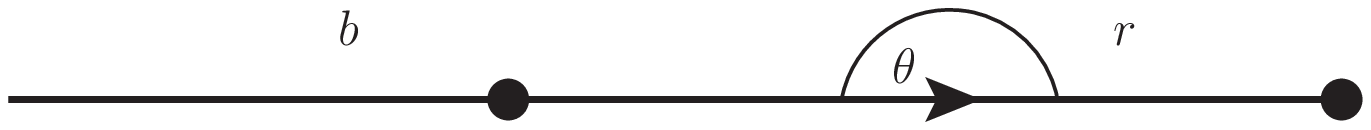}\hspace*{1in}
\includegraphics[angle=0,width=2.6in]{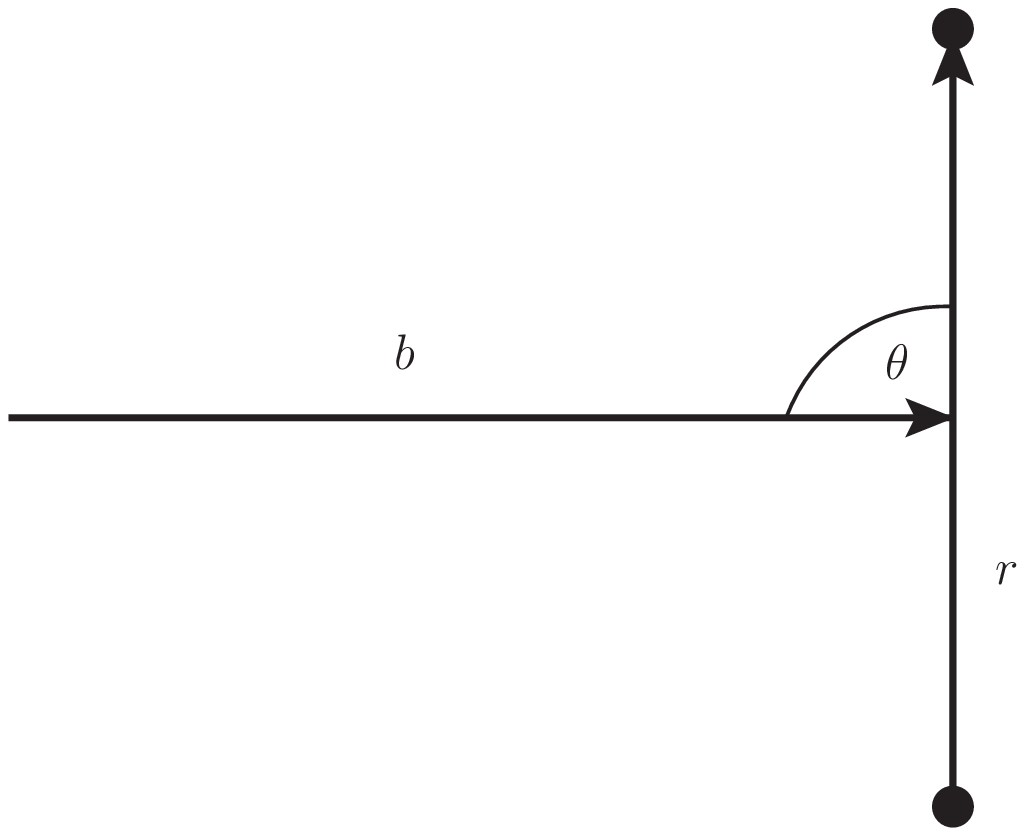}
\caption{Two configurations, 'aligned' and 'perpendicular', of the dipole orientation with respect to the impact parameter. 
\label{fig:dipolecartoon}}
\end{figure}
There also exists a nontrivial angular dependence which is most prominent in the  cases of large dipole size or impact parameter but for very specific configurations. In the case when the dipole size is much smaller or much larger than the impact parameter the solution does not depend much on the spatial orientation of the dipoles. On the other hand, for the case when the dipole size is twice as large as the impact parameter there exists strong angular dependence.   These effects are best illustrated  in Figs. \ref{fig:AngleCompR}(a,b)  and \ref{fig:AngleCompB}(a,b).  The scattering amplitude  as a function of  both dipole size and impact parameter for different choices of the angle $\theta$  is shown, where $\theta$ is  defined as the angle between the dipole and the impact parameter, as illustrated in Fig.~\ref{fig:dipolecartoon}.  The amplitude has a peak when at $r=2b$ for both orientations as is shown in both plots in  Figs. \ref{fig:AngleCompR}.  Note however, that the peak is distinctively sharper for the 'aligned' dipole configurations, when $\cos \theta = 1.0 \;  {\rm or} \; -1.0$, than for the 'perpendicular' configuration. This is also illustrated in  Fig.~\ref{fig:AnglePlots} where the dependence on the angle is shown. For values that are near the $r=2b$ point  there are enhancements at $\cos\theta = 1.0,-1.0$ and this is present in both kernels.  These effects can be seen in both plots in terms of dipole size and impact parameter.  
It is interesting to note that the peak is present in the case of scattering amplitude versus dipole size the peak even when $\cos\theta = 0.0$. On the other hand   such structure is absent for this configuration in the impact parameter profile with fixed dipole size.
It is also evident in Fig.~\ref{fig:AngleCompB}  that the amplitude is flat in impact parameter when the dipole size is much larger than $b$.
We will demonstrate that all these effects can be easily understood from conformal representation of the amplitude \ref{sec:conformal}.

\subsection{Saturation Scales}
\begin{figure}
\centering
\includegraphics[angle=270,width=0.49\textwidth]{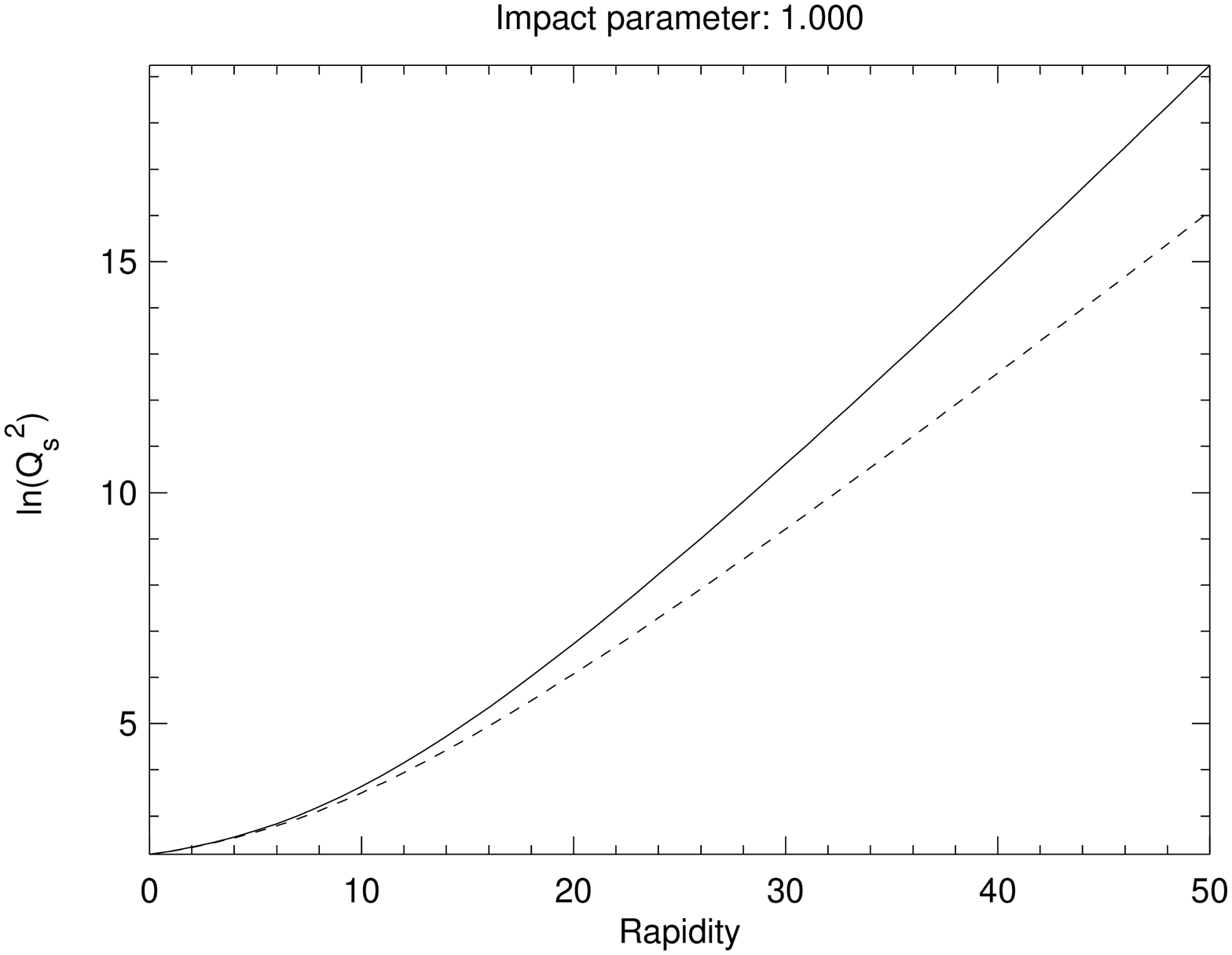}\hspace*{0.5cm}
\includegraphics[angle=270,width=0.49\textwidth]{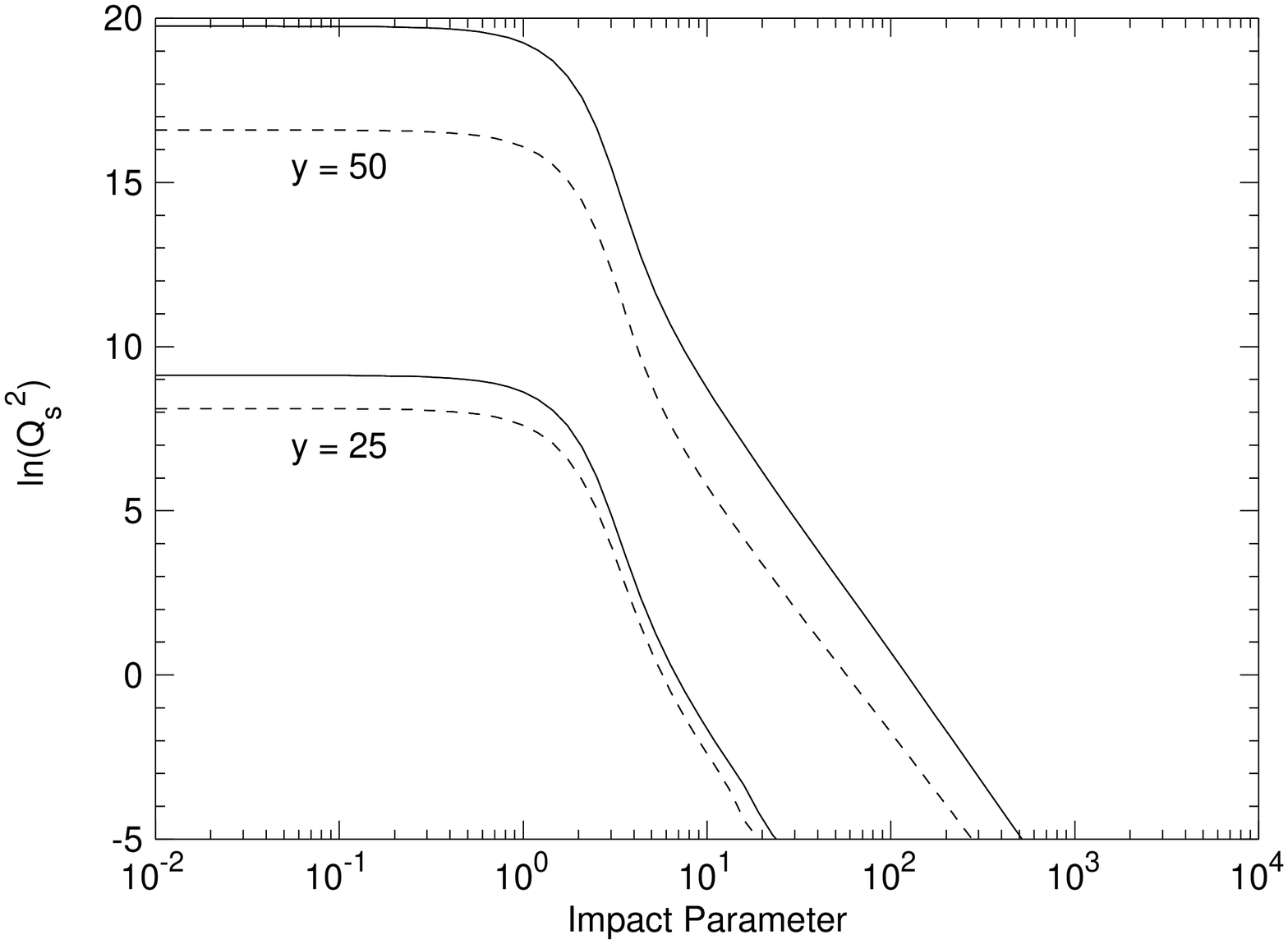}
\caption{Left: plot of the saturation scale as a function of rapidity.  The solid line corresponds to the LO simulation and the dashed line is the saturation scale of the Bessel kernel. The impact parameter is fixed at $b=1.0$. Right: the dependence of the saturation scale on impact parameter for two different values of rapidity. Solid line: LO kernel, dashed line Bessel kernel. Strong coupling is fixed at $\asb=0.1$.}
\label{fig:satscale}
\end{figure}

The saturation scale  in the impact parameter dependent scenario is again defined by the following equation
\begin{equation}
\langle N(r=1/Q_s,b,\theta,Y)\rangle \;  = \;  \kappa \; , 
\label{eq:satkappa}
\end{equation}
where $\kappa$ is a constant.  In all the following analysis we have set $\kappa = 0.5$.  It is important to note that, in this case the form of the amplitude admits two solutions to the above equation.
As is evident from  Fig.~\ref{fig:RdepUcor} one solution for the saturation scale is  for a larger dipole size and one for a smaller dipole size.  The saturation scale $Q_s$ always refers to the solution where the dipole size is smaller.  
We have found that  the slope in rapidity of the saturation scale $Q_s$ increases for low values of rapidities, then reaches an approximately constant value and for ultrahigh rapidities it starts to decrease.
The first effect is caused by the preasymptotic  contributions, the latter effect is caused by the finite size of the grid.  We have found that the effects of the grid can be neglected below the rapidities of order $\sim 60$. The saturation scale as a function of the rapidity is shown in  left plot in Fig.~\ref{fig:satscale}. The solid line shows the calculation in the case of the LO kernel and the dashed line is for the Bessel kernel. It is clear that, the dependence on the rapidity is exponential as expected for the computation with fixed value of the coupling. The numerical value of the exponent governing the rapidity dependence of the saturation scale, extracted in the LO kernel case is $\lambda_s = 4.4$, see (\ref{eq:satsmall}). The value of the exponent  extracted for the evolution with Bessel kernel
was found to be $\lambda_s = 3.6$. Clearly, the subleading effects of the modified kernel cannot be neglected  here, which has to be contrasted with the case without the impact parameter.

In the case when impact parameter is much larger than the inverse of the saturation scale the exponent in rapidity is independent of the impact parameter value. This means that  the saturation scale
has a factorized form
\be
Q_s^2(Y,b) = Q_0^2  \exp(\bar{\alpha}_s \lambda_s Y) S(b) \; .
\label{eq:satsmall}
\ee

This is demonstrated in Fig.~\ref{fig:satscale} where the small dipole saturation scale is shown as a function of the impact parameter for two different values of rapidity.  The power tail $1/b^4$ is clearly prominent. There is significant difference between the saturation scale from the Bessel function kernel and the LO kernel.

The second solution of the  equation (\ref{eq:satkappa}), which shall be called $Q_{sL}(Y,b)$, gives saturation scale at large dipole size.  The same analysis that was performed on $Q_s$ can be performed on $Q_{sL}$ with the parameterization of this  saturation scale taken to be

\begin{equation}
Q_{sL}^2 = Q_{0L}^2 e^{-\lambda_{sL} \bar{\alpha}_s Y} \; ,
\end{equation}

where once again $Q_{0L}$ is a normalization term and the minus sign in the exponent is because the evolution is now moving towards larger dipole sizes.
\begin{figure}
\includegraphics[angle=270,width=0.43\textwidth]{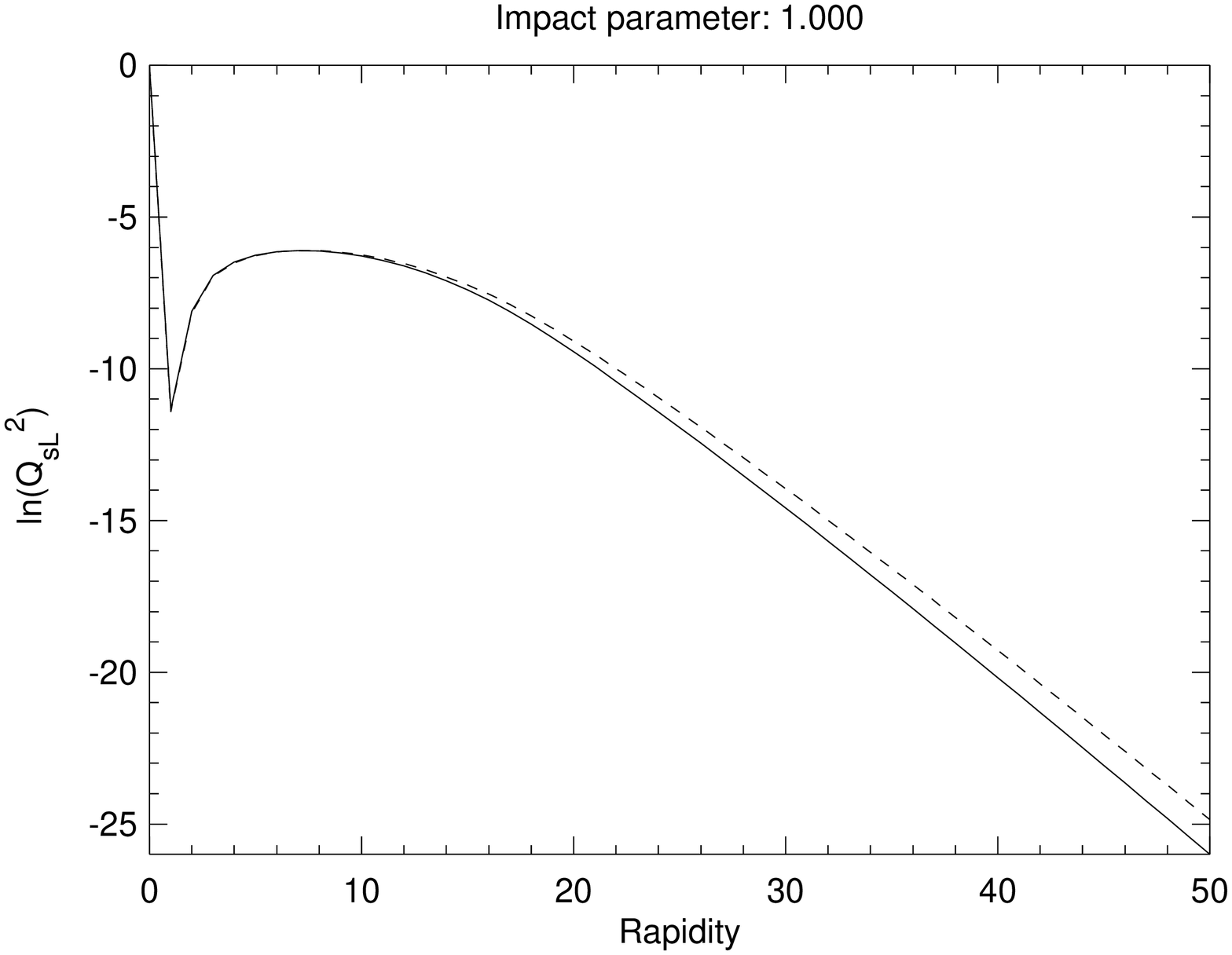}\hspace*{0.5cm}
\includegraphics[angle=270,width=0.43\textwidth]{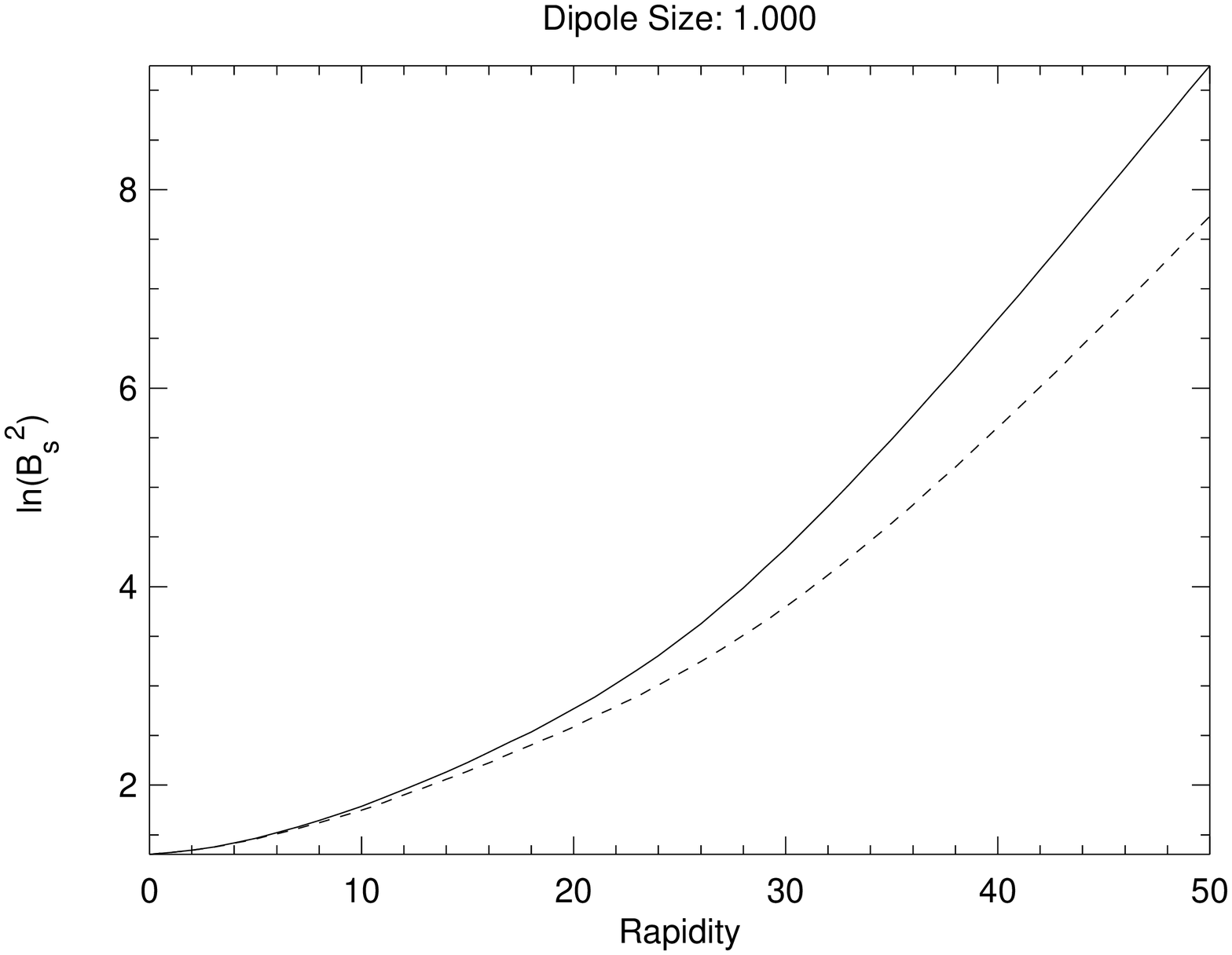}
\caption{Left: plot of the saturation scale for large dipole size  as a function of rapidity.  The solid line represents the result using LO kernel and the dashed line represents the result using the  Bessel kernel.   The impact parameter is fixed at $b=1.0$.
Right: plot of the black disc radius defined by Eq.~\ref{eq:bdr} for the LO kernel (solid line) and the modified kernel (dashed line).  The dipole size is fixed at $r=1.0$
\label{fig:LargeSat}}
\end{figure}

The extracted value for the LO kernel is $\lambda_{sL} = 6.0$ and for the Bessel function kernel $\lambda_{sL} = 5.6$.  The difference between these two exponents is now about 7\% and it can be seen on Fig. \ref{fig:LargeSat} that these curves are much closer than in Fig. \ref{fig:satscale}.  

The reason  for these effects can be understood by again 
 inspecting  the form of the Bessel kernel

\begin{equation}
Q^2_{01}\left[K_1^2(Q_{01}x_{02})+K_1^2(Q_{01}x_{12})-2K_1(Q_{01}x_{02})K_1(Q_{01}x_{12})\frac{x_{02}\cdot x_{12}}{|x_{02}||x_{12}|}\right] \;. 
\end{equation}

When $x_{01}$ is small then $Q_{01}$ is large and the cutoff is on dipoles such that $x_{02}>1/Q_{01}$ is very large.   This means that unless $x_{02}$ is very large  (which would also correspond to large $x_{12}$) then the modified kernel is very close the LO kernel.  It is this limitation of the phase space of the modified kernel that causes the dipole saturation scale for large dipoles to have a rather modest difference between the Bessel and LO case.  
At this point it is worth to note that the Bessel kernel does not exhaust all the kinematical effects. To be more precise, we would expect that the kinematical cuts are resulting in the kernel which is also conformally invariant. This will give cuts  on both small and large dipole sizes and further reduce the evolution speed.

One can also define from (\ref{eq:satkappa}) a scale which corresponds to the extension in impact parameter space. This scale is the radius of the black disc. This can be done by solving this equation for $b$ rather than for the dipole size.

We therefore define the black disc radius in impact parameter $B_s$ by solving the equation

\begin{equation}
\langle N(r,b=B_s,\theta)\rangle = \kappa \; ,
\label{eq:bdr}
\end{equation}
with respect to $b$, and
where once again $\kappa = 0.5$ is chosen.  We assume the exponential form for the behavior of the impact parameter radius as a function of rapidity
\begin{equation}
B_s^2 = B_{s0}^2 e^{\lambda_{Bs} \bar{\alpha}_s Y} \;.
\end{equation}

Here $B_{s0}$ is a normalization term and $\lambda_{Bs}$ is extracted from the numerical solution. We have found for the LO kernel $\lambda_{Bs} = 2.6$ and for the modified kernel $\lambda_{Bs} = 2.2$.  This is approximately half of the $\lambda_s$ in the case of the LO kernel. As argued before \cite{MishaRyskin, GolecBiernat:2003ym} this is due to the fact that the amplitude depends on one variable, which in the case of the configuration $b\gg r$ is proportional to $r/b^2$. This immediately means that the impact parameter dependence in rapidity is twice slower than the one of the dipole size.   The computation for the Bessel function kernel shows that it does not hold as closely in this case because in such case we do not have an exact conformal symmetry.
These properties will be explained  in more detail in the next section.
\begin{center}
\begin{table}
\begin{tabular}{| l || c | c | c |}
\hline
 & $\lambda_s$ & $\lambda_{sL}$ & $\lambda_{sB}$\\
\hline
LO Kernel (1) $\bar{\alpha}_s = 0.1$ & 4.4 & 6.0 & 2.6\\
LO Kernel (2) $\bar{\alpha}_s = 0.1$ & 4.4 & 5.8 & 2.6\\
Modified Kernel $\bar{\alpha}_s = 0.1$ & 3.6 & 5.8 & 2.2\\
\hline
LO Kernel $\bar{\alpha}_s = 0.2$ & 4.4 & 5.9 & 2.6\\
Modified Kernel $\bar{\alpha}_s = 0.2$ & 2.5 & 5.2 & 2.0\\
\hline
\end{tabular}
\caption{Summary of extracted saturation exponents for solutions with  impact parameter.
(1) means Glauber-Mueller initial conditions, (2) means Glauber -Mueller with the exponential cutoff on the large dipole sizes.}
\label{Table:LambdaSummary}
\end{table}
\end{center}

\begin{figure}[htb]
\subfigure[Impact parameter dependent amplitude  for the fixed coupling $\alpha_s = 0.1$ case as a function of the dipole size and for different values of the impact parameter.]{\label{fig:1}\includegraphics[angle=270,width=0.49\textwidth]{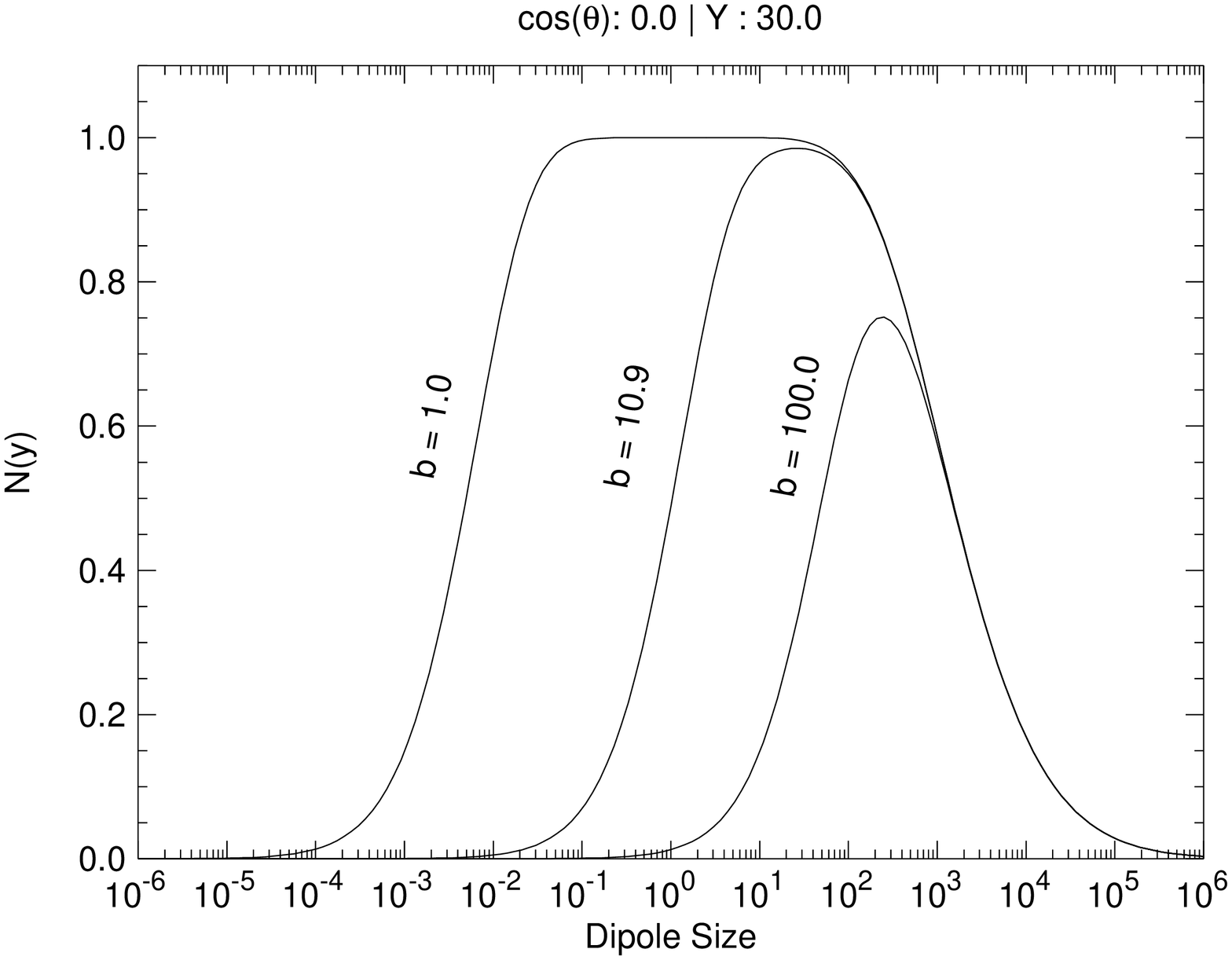}}
\subfigure[Saturation region in $(Y,\ln r)$ plane for different impact parameters.]{\label{fig:2}\includegraphics[angle=270,width=0.49\textwidth]{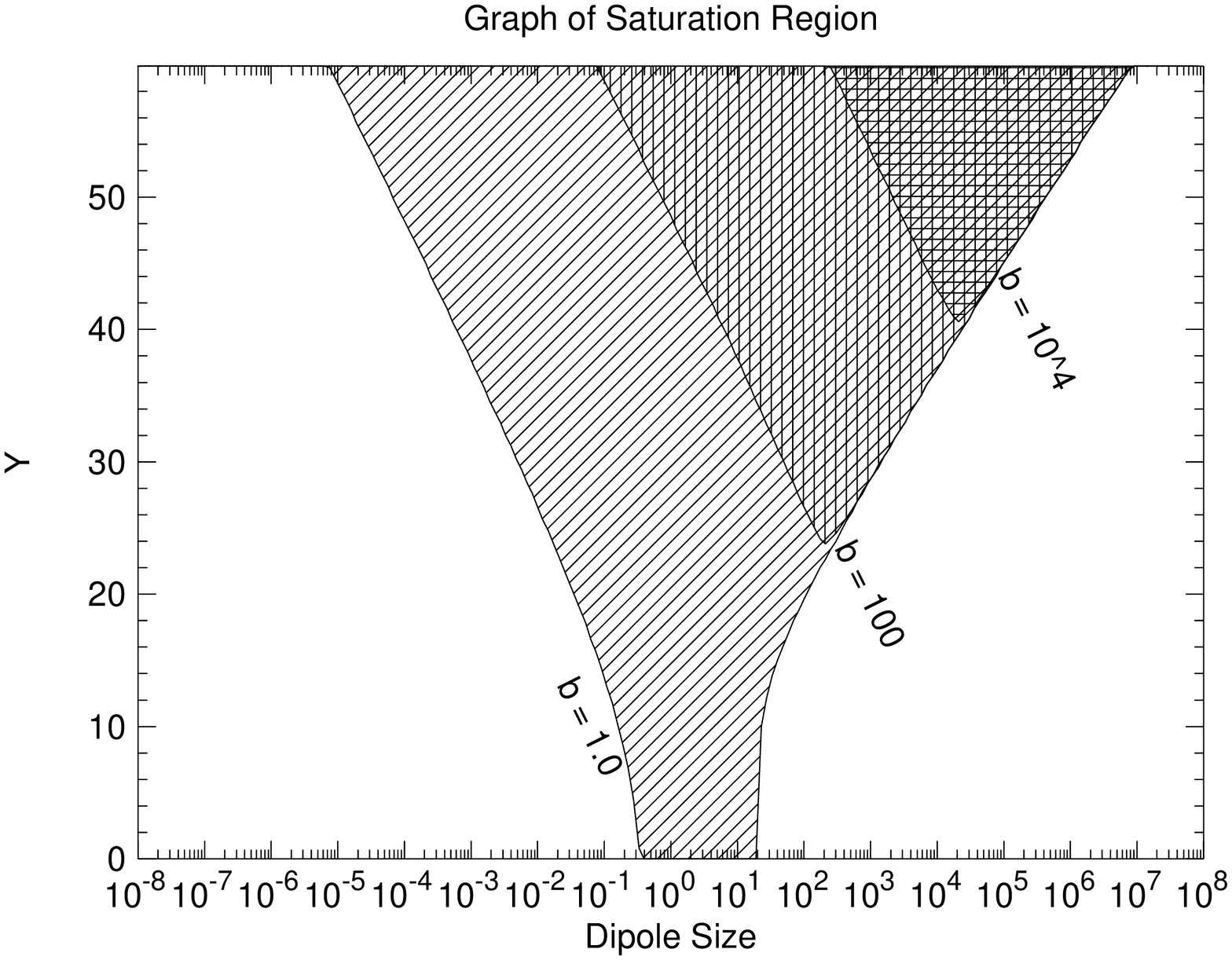}}
\caption{}
\label{fig:satregion}
\end{figure}

The simulations with different value of the fixed coupling were also performed. The results are summarized in table \ref{Table:LambdaSummary}. 
It can be seen that the exponents for the LO kernel do not depend on the value of the coupling,
which is the expected behavior as this is a fixed order calculation. On the other hand the exponents extracted for the modified kernel significantly differ exhibiting the nonlinearity in the coupling constant.
The exponents are further reduced with respect to the LO values which is due to the resummation of the subleading terms in $\ln 1/x$.

The dependence on the initial conditions was also tested. As an alternative, we have taken the second initial condition to be modified by the cutoff in the large dipole size
\be
N^{(0)}_{(2)}= N^{(0)} \exp(-r\mu) \; ,
\ee
where $\mu=1/5$.
The exponents are also shown in table  \ref{Table:LambdaSummary} (LO (2)). We observe a modest variation of the exponents with the change on the initial condition. The most significant change is for the large dipole saturation scale. This is to be expected as in this region the two initial conditions differ significantly.

In Fig.~\ref{fig:satregion}(a)  the amplitude as a function of the dipole size and various impact parameters is shown. It is interesting that for large dipole sizes the amplitude has the same front for all the impact parameters. This is related to the properties of the solutions stemming from the conformal symmetry, see Sec.~\ref{sec:conformal}. In Fig.~\ref{fig:satregion}(b) we show the saturation region for the solution with impact parameter. 
Unlike in the local case (without the impact parameter) here the saturation region has a 'V' shape in $(Y,\ln r)$ space, which is moving towards higher rapidities and larger dipole sizes as the impact parameter increases.  Different shaded areas correspond to three different impact parameters. Again, the common front for the different values of $b$ is clear. The distortion at lower rapidities and for small impact parameter stems from the initial conditions.

\subsection{Conformal representation and properties of the amplitude.}
\label{sec:conformal}

Most of the features  observed in the numerical solutions can be explained by using the conformal representation of the solution for the scattering amplitude. In general the representation can be shown to be of the form \cite{Lipatov:1985uk}
\be
F(\rb_0,\rb,\bb;Y)  \; = \; \int_{c-i\infty}^{c+\infty}\frac{d\omega}{2\pi i} \,  \exp(\omega Y) \,  F(\rb_0,\rb,\bb;\omega) \; ,
\ee
with
\begin{multline}
F(\rb_0,\rb,\bb;\omega) = \sum_{n=-\infty}^{+\infty} \int_{-\infty}^{\infty} d\nu \int d^2 \wb \,\left(\nu^2+\frac{n^2}{4}\right) \,  \frac{{\cal F}(\nu,n,\alpha_s;\omega)}{\left[\nu^2+(\frac{n-1}{2})^2\right]\left[\nu^2+(\frac{n+1}{2})^2\right]} \, \times \\
\times \,  {E^{n,\nu}}^*\left(\frac{\rb_0}{2}-\wb,-\frac{\rb_0}{2}-\wb\right) \, E^{n,\nu}\left(\frac{\rb}{2}+\bb-\wb,-\frac{\rb}{2}+\bb-\wb\right) \; ,
\label{eq:conformalrep}
\end{multline}
\label{eq:oniumonium}
where $\rb_0,\rb$ are the transverse sizes of two scattering objects (for example onia), $\bb$
is their relative impact parameter.  

The conformal eigenfunctions are defined as
\be
E^{n,\nu}(\rho_{10},\rho_{20}) \; = \; (-1)^n \left( \frac{\rho_{12}}{\rho_{10}\rho_{20}} \right)^h \left( \frac{\rho_{12}^*}{\rho_{10}^*\rho_{20}^*} \right)^{\bar{h}} \; ,
\label{eq:econf}
\ee
where complex notation for the two dimensional vectors  $(\rho_x,\rho_y)$ has been used
$$\rho=\rho_x+i \rho_y, \rho^*=\rho_x-i \rho_y\;,$$
and where the conformal weights are
$$h=\frac{1-n}{2}+i\nu, \; \; \; \; \bar{h}=\frac{1-n}{2}+i\nu\; .$$
Function ${\cal F}(\nu,n,\alpha_s;\omega)$ contains the details of the dynamics. For the case of evolution with linear BFKL, the form of it is well known
$$
{\cal F}^{\rm BFKL}(\nu,n,\alpha_s;\omega) \, = \, \frac{1}{\omega-\chi(n,\nu)} \; .
$$
with $$\chi(n,\nu) \; =  \; 2\Psi(1) - \Psi(\frac{1+|n|}{2}+i\nu) -\Psi(\frac{1+|n|}{2}-i\nu) \; ,$$
being the LO BFKL kernel eigenvalue.
In the case of the nonlinear equation the exact form of the function ${\cal F}(\nu,n,\alpha_s;\omega)$ is unknown.
The origins of the peaks in the amplitude can be understood by analyzing the transverse structure encoded in functions $E^{n,\nu}$.  We fix $\rb_0$ and investigate the dependence on $\rb$ from the
transverse integral. We switch from the vector notation to the complex notation for the arguments of the $E$ functions. Using the explicit expression (\ref{eq:econf}) we obtain

\begin{multline}
\int d^2 w    E^{n,\nu *}(\frac{r_0}{2}-w,-\frac{r_0}{2}-w) \, E^{n,\nu}(\frac{r}{2}+b-w,-\frac{r}{2}+b-w) = \\
= \int d^2 w E^{n,\nu *}(\frac{r_0}{2}-w,-\frac{r_0}{2}-w) (-1)^n \left(\frac{r}{(b+\frac{r}{2}-w)(b-\frac{r}{2}-w)}\right)^h\left(\frac{{r}^*}{(b+\frac{r}{2}-w)^* (b-\frac{r}{2}-w)^*}\right)^{\bar{h}} \; ,
\end{multline}
where we switched to the complex notation for the arguments. The biggest contribution comes from the region of $w\simeq 0$. For our purposes it is also enough to take $n=0$.
In this region the integrand has the approximate form
\be
E^{n,\nu *}(\frac{r_0}{2},-\frac{r_0}{2}) \left(\frac{|r|^2}{[b^2-(\frac{r}{2})^2][(b^*)^{2}-(\frac{r^*}{2})^{2}]} \right)^{1/2+i\nu} \; .
\ee 
Using $b=|b|e^{i\theta_b},r=|r|e^{i\theta_r}$ and $\Delta \theta=\theta_r-\theta_b$ we have that
\be
E^{n,\nu *}(\frac{r_0}{2},-\frac{r_0}{2}) \left(\frac{|r|^2}{[|b|^2-e^{2i\Delta \theta}(\frac{|r|}{2})^2][|b|^{2}-e^{-2i\Delta \theta}(\frac{|r|}{2})^{2}]} \right)^{1/2+i\nu} \; .
\ee 
It is immediately clear that there will be  angular dependence for the $b \simeq \frac{r}{2}$ case with the configurations  of 'aligned' dipoles $\Delta \theta = 0,\pi$ giving largest contributions.  In the case of the 'perpendicular' orientation of dipoles with respect to the impact parameter $\theta=\pi/2,-\pi/2$, the expression reduces to 
\be
E^{n,\nu *}(\frac{r_0}{2},-\frac{r_0}{2}) \left(\frac{|r|}{|b|^2+(\frac{|r|}{2})^2} \right)^{1+2 i\nu} \; .
\ee 
This structure is responsible for the presence of the peak in the amplitude in the case when the $b$ is fixed and $r$ varied, and the absence of the peak in the case when $r$ is fixed and $b$ varied, for $\Delta \theta=\pi/2,-\pi/2$. This corresponds to the situations in right hand plots in Figs.~\ref{fig:AngleCompR} and \ref{fig:AngleCompB}
correspondingly.

An expression for the saturation scale dependent on the impact parameter can be derived using the method in \cite{Mueller:2002zm}. To this aim one needs to take the Mellin representation for the solution to the linear equation and apply the absorptive boundary.  
The integral over the transverse variable  can be performed using the representation \cite{Lipatov:1985uk}
\begin{multline}
\int d^2 {\rho_0} E^{n,\nu}(\rho_{10},\rho_{20}) E^{n,\nu*}(\rho_{1'0},\rho_{2'0}) \; = \; 
c_1 x^{h}x^{*\bar{h}} F(h,h,2h;x)\,F(\bar{h},\bar{h},2\bar{h};x^*) \\
+ c_2 x^{1-h}x^{*1-\bar{h}} F(1-h,1-h,2-2h;x) F(1-\bar{h},1-\bar{h},2-2\bar{h};x^*)  \;,
\end{multline}
where 
\be
x = \frac{\rho_{12}\rho_{1'2'}}{\rho_{11'}\rho_{22'}} \; ,
\label{eq:anratio}
\ee
is the anharmonic ratio and $F$ are the hypergeometric functions. To obtain the saturation scale we take $n=0$, and expand around $x\simeq 0$. This simplifies the above expressions as in this limit $F(h,h,2h;x) \sim 1$ and the whole dependence on $x$ comes through factors $x^{h}x^{*\bar{h}}$. In the case when the impact parameter $b$ is much larger than the dipole sizes, $b \gg r,r_0$, one has 
$$
|x| \simeq \frac{r r_0}{b^2} \; .
$$
Putting everything together, the scattering the amplitude in the linear evolution case reduces to 
\cite{Navelet:1997tx,Hatta:2007fg}

\begin{equation}
F(\rb_0,\rb,\bb;Y) = -\frac{ \alpha_s^2 r_0 r}{b^2}  \int_{-\infty}^{\infty} \frac{d\nu}{2\pi}   \,  \frac{i\nu}{\left[\nu^2+\frac{1}{4}\right]^2}   \exp\left({\bar{\alpha}_s \chi(\lambda) Y+2 i \nu \ln\frac{b^2}{r_0 r}}\right) \; .
\label{eq:NavaletWallonAmp1}
\end{equation}
Here, we have taken into account the contribution from only zero conformal spin.
Using $\lambda=1/2+i\nu$ the above expression can be recast into

\begin{equation}
F(\rb_0,\rb,\bb;Y) = - \alpha_s^2  \int_{\frac{1}{2}-i\infty}^{\frac{1}{2}+i\infty} \frac{d\lambda}{2\pi}   \,  2 (\lambda-1/2) \,   \exp\left({\bar{\alpha}_s \chi(\lambda) Y -(1-\lambda) \ln\left(\frac{b^2}{r_0 r}\right)^2}\right) \; ,
\label{eq:NavaletWallonAmp2}
\end{equation}
where the prefactor in (\ref{eq:NavaletWallonAmp1}) has been expanded around $\nu=0$.
Taking the saddle point condition and the condition that the exponent vanishes at the saddle point which is the requirement on the saturation boundary one arrives at two conditions for this line (noted by a subscript 0).

\begin{eqnarray}
\bar{\alpha}_s Y \chi'(\lambda_0)+\ln(b^2/(r_0 r))^2 &=& 0 \; ,
\\\bar{\alpha}_s Y \chi(\lambda_0)-(1-\lambda_0)\ln(b^2/(r_0 r))^2 &=& 0 \; .
\end{eqnarray}

These equations can be solved to yield the saturation scale but it was found that
one can include further corrections \cite{Mueller:2002zm}.
We can obtain these corrections by using the solution to the saddle point equation to find $\lambda_c$, and use this to evaluate the prefactor.  The resulting modified equations are then

\begin{eqnarray}
\bar{\alpha}_s Y \chi'(\lambda_c)+ \ln(b^2/(r_0 r))^2 &=& 0 \; ,
\\\bar{\alpha}_s Y \chi(\lambda_c)-(1-\lambda_c) \ln(b^2/(r_0 r))^2 &=& \frac {3}{2} \ln[ \bar{\alpha}_sY\chi''(\lambda_c)] \; .
\end{eqnarray}

By keeping one of the dipole sizes fixed, say $r_0$, we can solve for $r$ to get the saturation line
\begin{equation}
Q_{c,1}^2(r_0,b;Y) = \frac{r_0^2}{b^4} \frac{e^{\frac{\bar{\alpha}_s Y \chi(\lambda_c)}{(1-\lambda_c)}}}{  \, [ \bar{\alpha}_sY\chi''(\lambda_c)]^{\frac{3}{2(1-\lambda_c)}}} \; .
\end{equation}

For large $\bar{\alpha}_s Y$ the $\lambda_c$  approaches  $\lambda_0$ value, with $\lambda_0 = 0.37$.   The saturation scale has $1/b^4$ dependence which comes automatically from conformal symmetry.
One can solve the above equation for $b$ and keep $r_0,r$ fixed which yields

\begin{equation}
B_s^2(r_0,r;Y) = r_0 r\frac{e^{\frac{\bar{\alpha}_s Y \chi(\lambda_c)}{2(1-\lambda_c)}}}{ [ \bar{\alpha}_sY\chi''(\lambda_c)]^{\frac{3}{4(1-\lambda_c)}}} \; .
\end{equation}
This is the rate of the expansion of the radius in impact parameter space. Note that, the speed of the expansion is governed by the exponent which is half that of the saturation scale and the dependence on the dipole size is linear. This is also found in the numerical solution.
For the large dipole sizes $r \gg r_0,b$ the anharmonic ratio reduces to
$$
|x| \simeq \frac{4r_0}{r} \; .
$$
Following the same scheme one obtains for the saturation scale
\begin{equation}
Q_{c,2}^2(r_0,b;Y) =  \frac{e^{\frac{\bar{\alpha}_s Y \chi(\lambda_c)}{(1-\lambda_c)}}}{  \, r_0^2 [ \bar{\alpha}_sY\chi''(\lambda_c)]^{\frac{3}{2(1-\lambda_c)}}} \; .
\end{equation}
The saturation scale for large dipole sizes is independent of the impact parameter $b$.
This is also found in the solution, as is clear in Fig.~\ref{fig:satregion}. Therefore the 'V' shape of the saturation region is a consequence of the conformal symmetry of LO kernel.
From the above considerations one can see that the rapidity behavior of both saturation scales is identical for large and small dipoles. We found  that the two exponents differ somewhat, see Table \ref{Table:LambdaSummary}.  Most probably this is due to the initial conditions which are asymmetric in both small and large dipole sizes. However, a more detailed analysis is needed to confirm this effect.

In general we see that, both saturation scales, are in fact originating from one saturation scale due to the fact that the solution is expressed in terms of the anharmonic ratio. 
\subsection{Dipole cross section and black disc radius}

By integrating the amplitude over the impact parameter  the dipole cross section is obtained as a function of the dipole size and rapidity.  Despite the fact that the amplitude is bound and never exceeds unity, the dipole cross section can still increase very fast due to the fact that the amplitude has power tails in impact parameter, see discussion in \cite{Kovner:2002yt,Kovner:2001bh}. We thus expect the power like growth of the dipole cross section with the energy, or exponential with rapidity.

The  dipole cross section is defined as an integral over $b$ of the amplitude

\begin{equation}
\sigma(r,Y) \, = \, 2\int d^2 {\bb} \, N(\rb,\bb,Y) \; .
\end{equation}

In what follows, we will investigate the part of the dipole cross section which is coming from the black disc regime. To be precise, we integrate the amplitude over the values which are close to unity.  This is once again performed by constraining the amplitude through the equation (\ref{eq:bdr}).

The black disc part of the cross section is therefore defined as

\begin{equation}
\sigma_{\rm BD}(r,Y) \, = \, 2\int d^2{\bb} \,  N(\rb,\bb,Y) \Theta[N(\rb,\bb,Y) - \kappa] \; \approx \;  2 \pi R^2_{\rm BD}(r,Y) \; ,
\label{eq:BlackDisk}
\end{equation}

\begin{figure}
\centering
\subfigure[$r=0.11$]{\label{fig:CSBD1}\includegraphics[angle=270,width=0.3\textwidth]{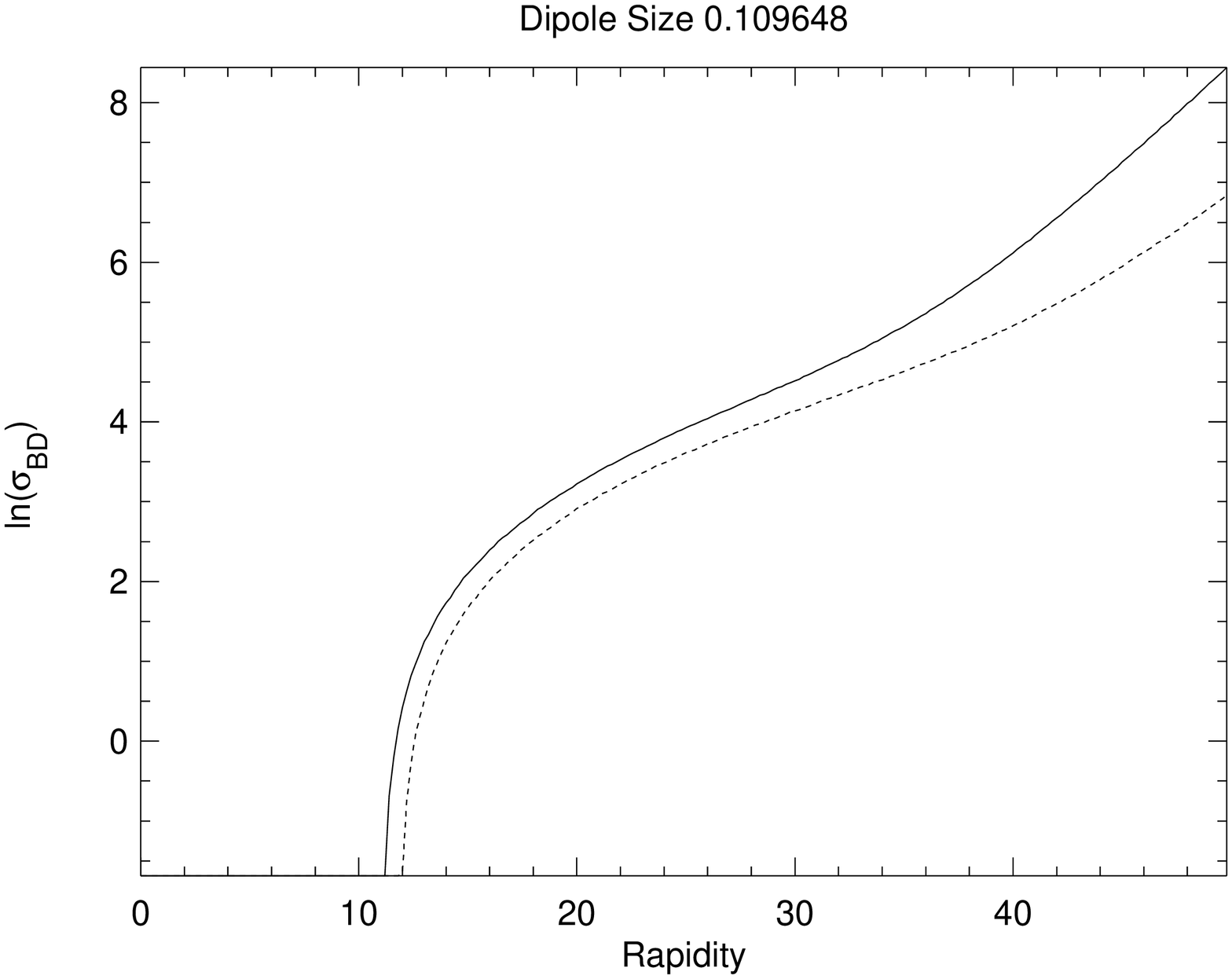}}
\subfigure[$r=1.0$]{\label{fig:CSBD2}\includegraphics[angle=270,width=0.3\textwidth]{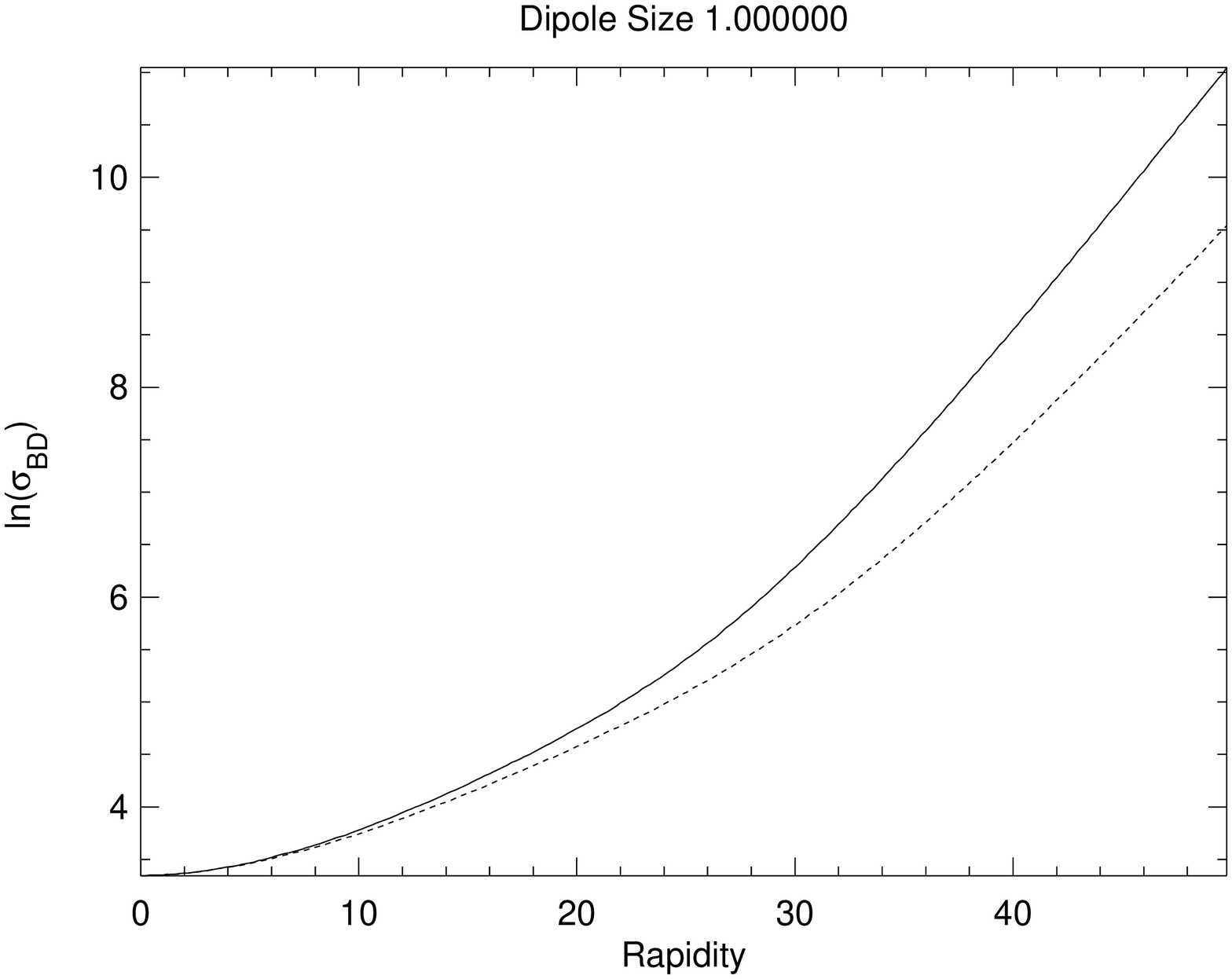}}
\subfigure[$r=11.0$]{\label{fig:CSBD3}\includegraphics[angle=270,width=0.3\textwidth]{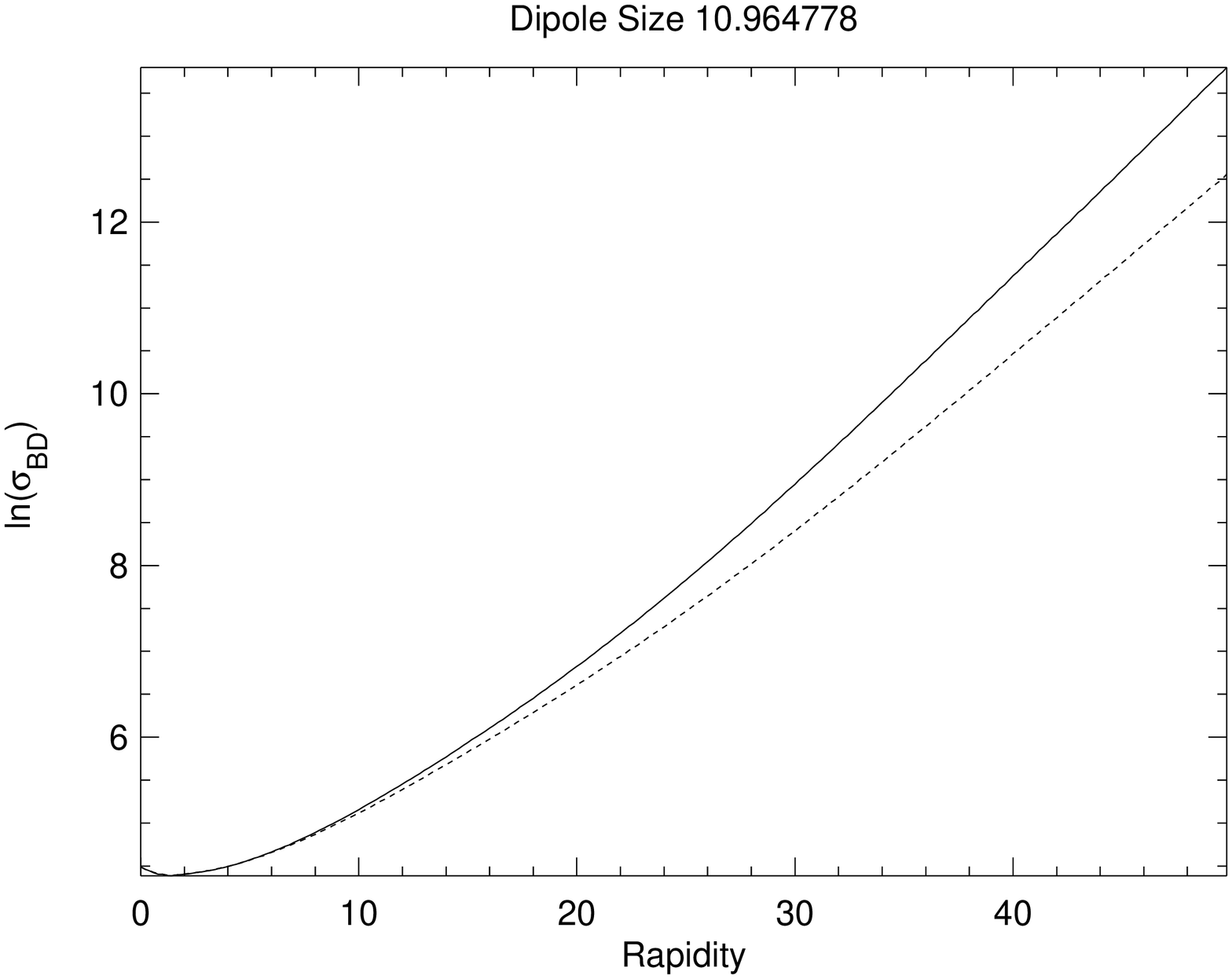}}
\caption{Graphs of the natural logarithm of the black disc cross section versus rapidity at various fixed dipole sizes.  The dashed line represents the solutions obtained with Bessel kernel while the solid line represents the solutions with the  LO kernel.}
\label{fig:CrossSectionBD}
\end{figure}

This black disc cross section is plotted  in Fig. \ref{fig:CrossSectionBD} and it can be seen that the slope of the black disc cross section in rapidity reaches a constant value at large rapidities.  One can parametrize $R_{BD}$ as

\begin{equation}
R_{\rm BD}^2(r,Y) \; = \; R_{\rm BD}^{(0)2} e^{\lambda_{\rm BD} \bar{\alpha}_s Y} \; ,
\end{equation}

where $R_{\rm BD}^{(0)}$ is a normalization constant and $\lambda_{BD}$ is extracted from the numerical solutions in the regime where it is approximately constant.  These extracted values for the solutions with two kernels as well as various values of $\bar{\alpha}_s$ are found in Table \ref{Table:LambdaBDSummary}.    The table also shows that with changing $\bar{\alpha}_s$ the exponent is relatively constant for the LO kernel, once again nonlinearities in the exponent appear for the Bessel kernel.  Reported exponents are averaged from values of dipole size $r = 10^{-1}\rightarrow 10^1$ because $\lambda_{\rm BD}$ does vary slightly with dipole size.  

\begin{center}
\begin{table}
\begin{tabular}{| l || c |}
\hline
 & $\lambda_{BD}$ \\
\hline
LO Kernel $\bar{\alpha}_s = 0.1$ & 2.4  \\
Modified Kernel $\bar{\alpha}_s = 0.1$ & 2.0  \\
\hline
LO Kernel $\bar{\alpha}_s = 0.2$ & 2.6 \\
Modified Kernel $\bar{\alpha}_s = 0.2$ & 1.6\\
\hline
\end{tabular}
\caption{Extracted exponents governing the behavior of  the black disc cross section.}
\label{Table:LambdaBDSummary}
\end{table}
\end{center}

\section{Including the  running coupling }
\label{sec:runcoupling}

Turning now to the running of the QCD coupling, $\bar{\alpha}_s$ is taken to be
${\alpha}_s(r) = \frac{1}{b \ln(\frac{1}{r^2 \lambda^2})}
$where $b = \frac{33-2n_f}{12\pi}$, $n_f$ is the number of active flavors and $\lambda = 0.246 \; {\rm GeV}$ is used.   In the infra-red regime,   the coupling was frozen when $r>r_{\rm cut}$ which is defined as  $\bar{\alpha}_s(r_{cut}) = 0.3$.  
As it is well known the BK equation without the impact parameter is not very sensitive to the way the coupling is regularized.  This is because the amplitude is saturated for all the large values of the dipole size from the inverse of the saturation scale to infinity.
   In the case with impact parameter however, there are contributions from the large dipole regime which spoil this self-regularizing behavior. In this case there is a large sensitivity to the regularization scenario for the running coupling.

There are two different schemes  for including the running coupling in  the BK equation \cite{Balitsky:2006wa}, \cite{Kovchegov:2006vj}.  In addition to these two scenarios we will use also the so-called parent dipole scheme, where the coupling depends on the size of the external dipole, that is $x_{01}$. This scheme   is convenient to use with the Bessel function kernel.    We have also evaluated the solutions using the prescription proposed in \cite{Balitsky:2006wa}
\begin{equation}
K^{Bal}(x_{01},x_{02}) = \frac{N_c \alpha_s(x_{01}^2)}{2\pi^2}\left[\frac{x_{01}^2}{x_{02}^2x_{12}^2} + \frac{1}{x_{02}^2}\left(\frac{\alpha_s(x_{02}^2)}{\alpha_s(x_{12}^2)} - 1\right) + \frac{1}{x_{12}^2}\left(\frac{\alpha_s(x_{12}^2)}{\alpha_s(x_{02}^2)} - 1\right)\right] \; .
\label{eq:Balitsky}
\end{equation}
Since it is not clear at the moment how to use this scheme with the Bessel function kernel we will use it only with the LO kernel.
 The scheme dependence between two prescriptions \cite{Balitsky:2006wa}, \cite{Kovchegov:2006vj} originates from the choice of the subtraction
point.  
The scheme by \cite{Kovchegov:2006vj} was shown to agree with the scheme \cite{Balitsky:2006wa} by the calculation of the appropriate subtraction corrections.
In this paper we have not evaluated the scheme \cite{Kovchegov:2006vj}, as we have found that in order to achieve the desired accuracy for the solution with impact parameter within this scheme takes considerably longer time.

\begin{figure}
\centering
\subfigure[ Solid line: LO kernel with fixed $\bar{\alpha}_s = 0.1$; dashed line: LO kernel with $\bar{\alpha}_s = \bar{\alpha}_s(r)$ ]{\label{fig:RNRcomp1d}\includegraphics[angle=270,width=0.49\textwidth]{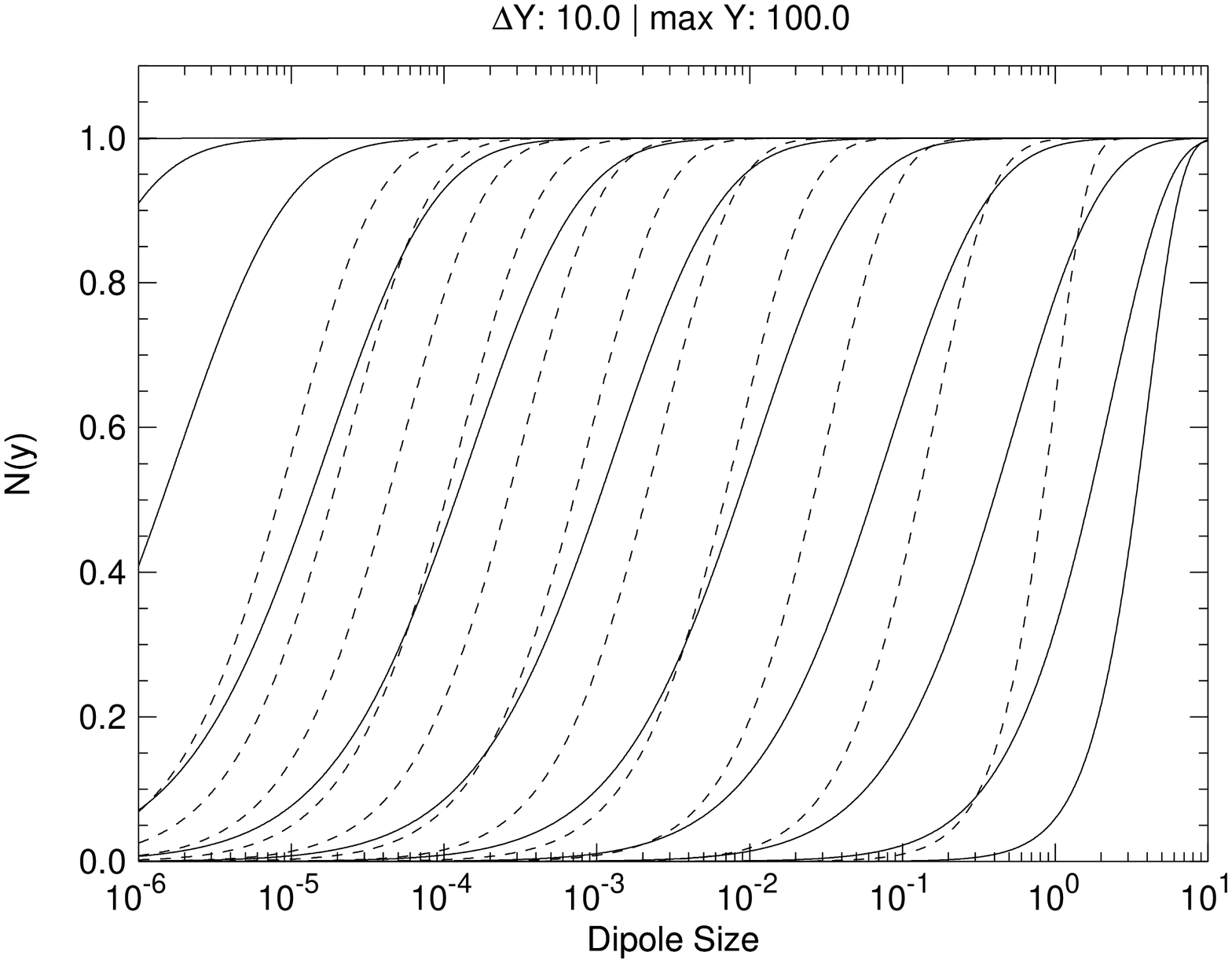}}
\subfigure[ Solid line: LO kernel with $\bar{\alpha}_s = \bar{\alpha}_s(r)$;  dashed line: modified kernel $\bar{\alpha}_s = \bar{\alpha}_s(r)$ ]{\label{fig:Rcomp1d}\includegraphics[angle=270,width=0.49\textwidth]{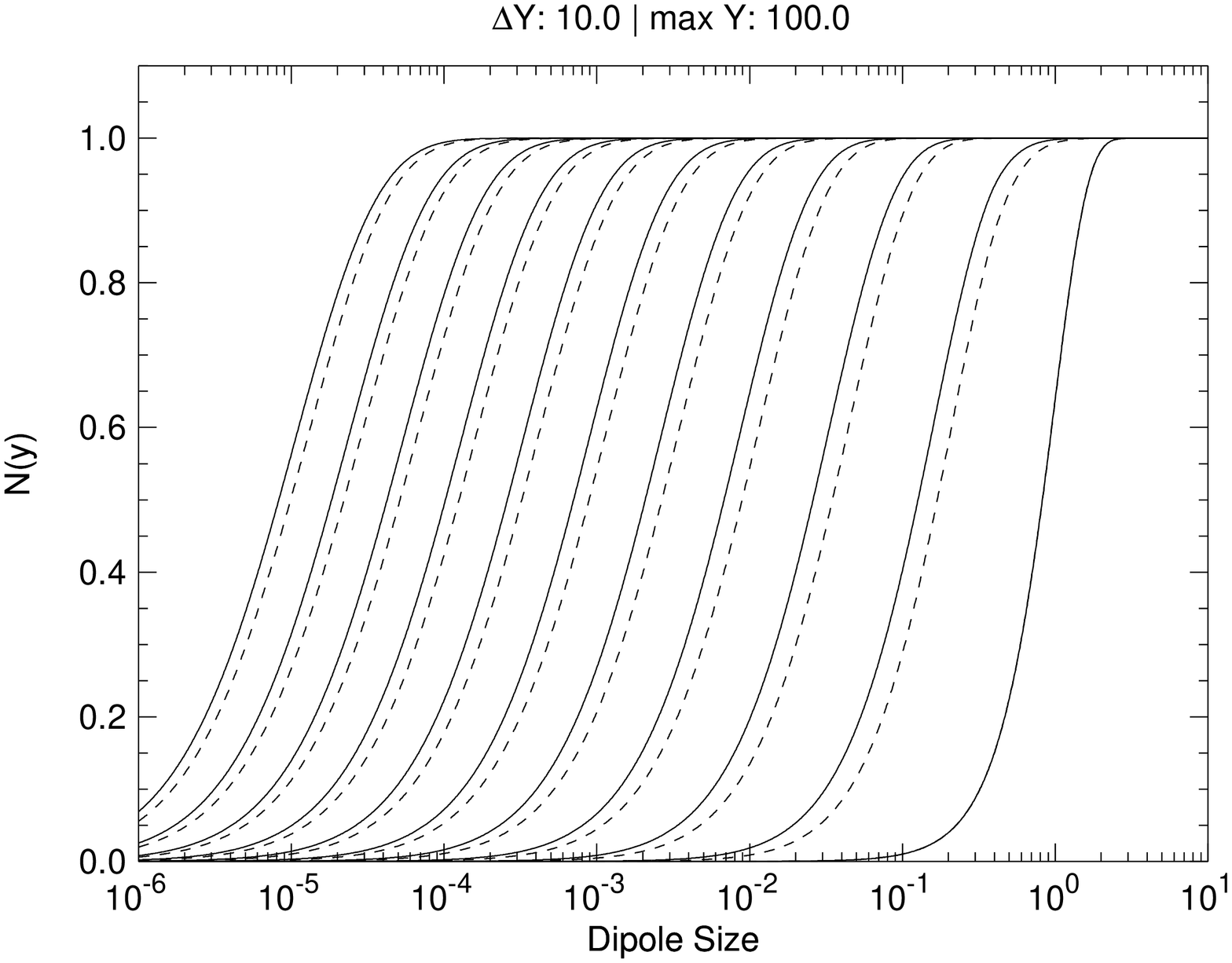}}
\caption{Graphs of the scattering amplitude versus dipole size for the case without impact parameter.  The left graph illustrates how much slower propagation is due to the running of the coupling and the right graph shows the modified and LO kernels compared to each other in the case of the running coupling.  Each line corresponds to the rapidity increasing in rapidity in intervals of $\Delta Y=10$ up to $Y=100$.}
\label{fig:Run1d}
\end{figure}

\begin{figure}
\includegraphics[angle=270,width=4.5in]{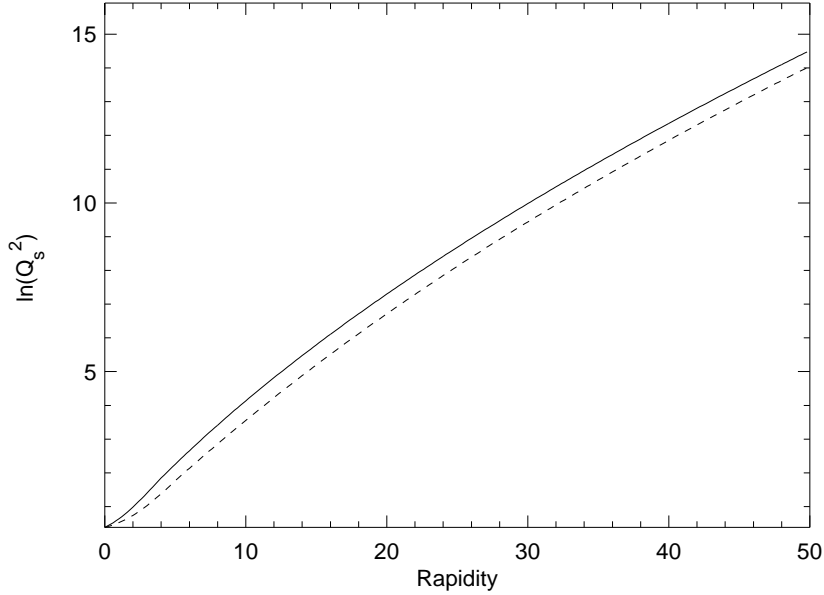}
\caption{Graph of the saturation scale of the LO kernel (solid line) and the modified kernel (dashed line) with running coupling and no impact parameter dependence. \label{fig:Run1dSat}}
\end{figure}

We first shall show the results with the running coupling without the impact parameter.  The running of the coupling has the effect of slowing down the evolution of the scattering amplitude as seen in Fig.\ref{fig:Run1d}.  The difference between the LO and the modified kernel with running coupling is rather small.  This can also be seen in Fig.\ref{fig:Run1dSat} which shows the saturation scale of the two kernels with running coupling which are extremely close to each other.
The dependence on the saturation scale with respect to the rapidity is as in \cite{Munier:2003sj} 
\begin{equation}
Q_s^2 = \lambda^2 \exp{\left[\left(\frac{2\chi(\gamma_c)}{b\gamma_c}Y\right)^{\frac{1}{2}}+\frac{3}{4} \xi_1 \left(\frac{\chi^{''}(\gamma_c)}{2b\gamma_c\chi(\gamma_c)}\right)^{1/3}Y^{\frac{1}{6}}\right]} \; ,
\label{eq:runsat}
\end{equation}
where $\xi_1=-2.338$.
Here the second term involving $Y^{\frac{1}{6}}$ is numerically non-negligible for the rapidities we consider. In terms of numbers the coefficients above give $Q_s^2 = \lambda^2 e^{3.6Y^{\frac{1}{2}}-5.4Y^{\frac{1}{6}}}
$   We have found that   the LO saturation scale  with running coupling and the parent  dipole size prescription is $Q_s^2 = e^{3.4 Y^{\frac{1}{2}}-4.8Y^{\frac{1}{6}}}$ which is very similar to the one given by the analytical value.  The running coupling with prescription (\ref{eq:Balitsky}) has also been run and found to have a fit of $Q_s^2 = e^{3.4 Y^{\frac{1}{2}}-5.7Y^{\frac{1}{6}}}$ which is closer to the value given by (\ref{eq:runsat}).

\begin{figure}
\centering
\subfigure[]{\label{fig:RunComp1}\includegraphics[angle=270,width=0.45\textwidth]{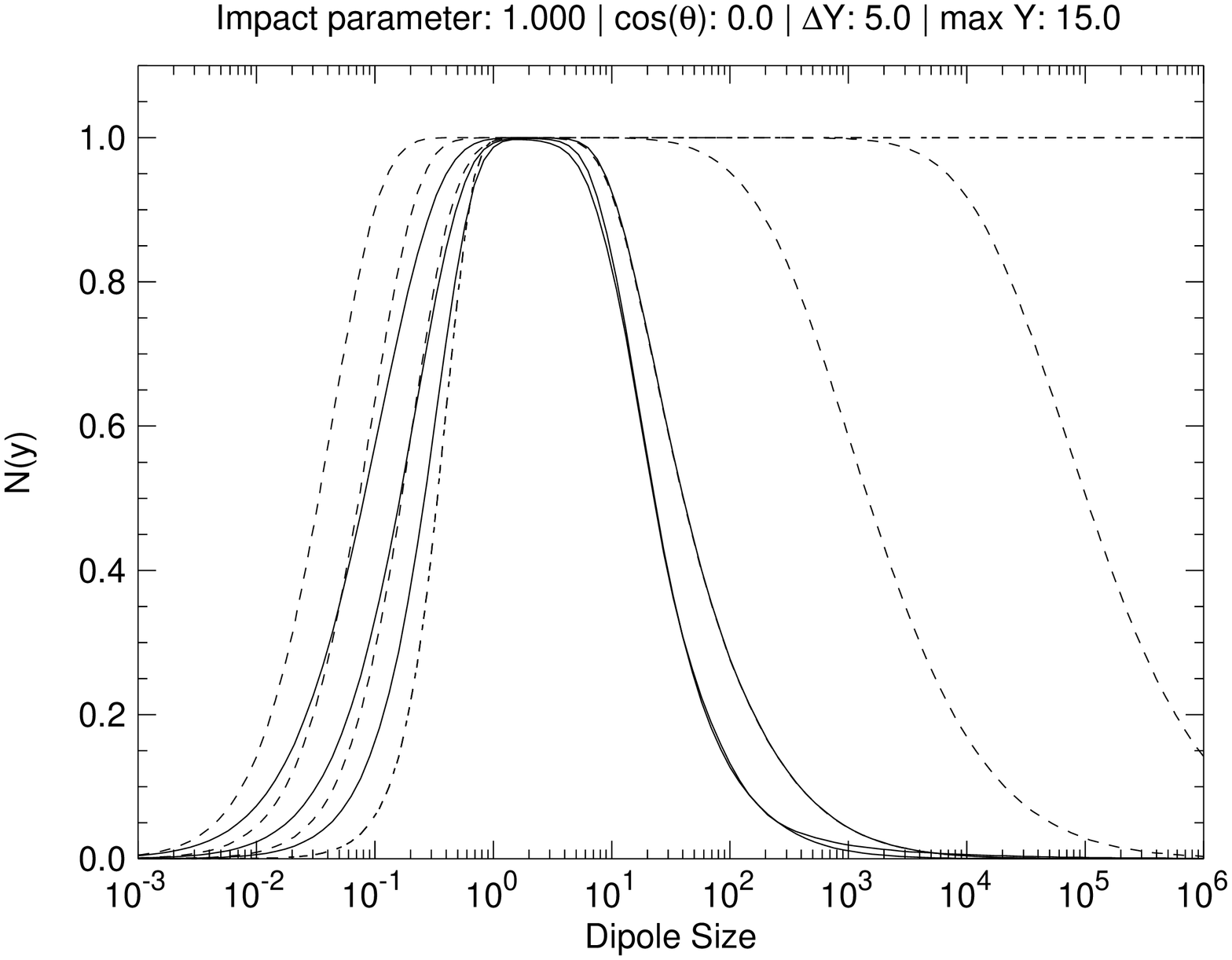}}
\subfigure[]{\label{fig:RunComp2}\includegraphics[angle=270,width=0.43\textwidth]{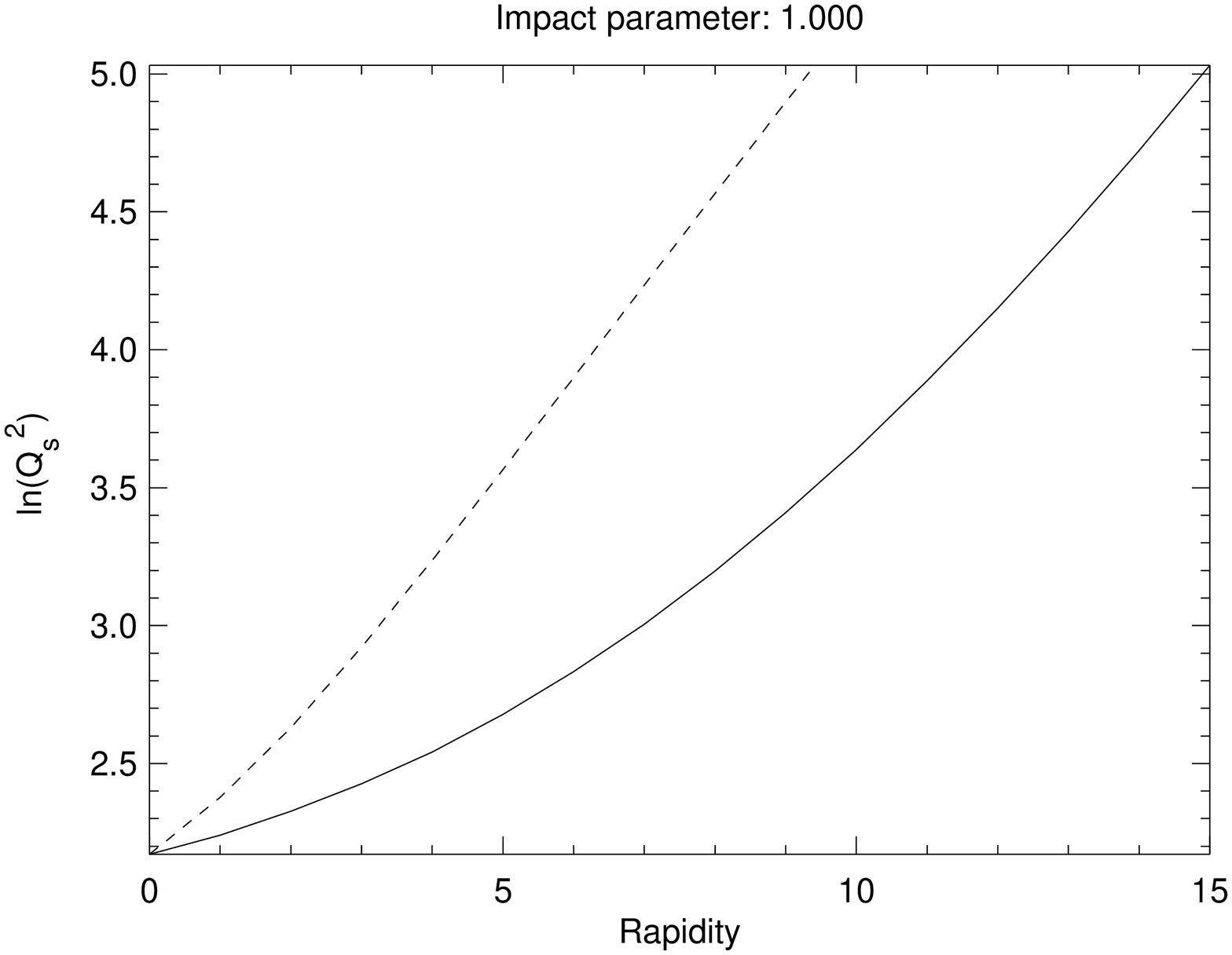}}
\caption{
Left (a) dipole scattering amplitude as a function of the dipole size for fixed impact parameter and angle.
Solid lines: fixed coupling with $\asb=0.1$, dashed lines: running coupling with the parent dipole scheme.
Right (b): saturation scale as function of rapidity for the LO kernel   the fixed coupling  $\asb=0.1$ (solid line) and running coupling (dashed line) case.}
\label{fig:RunCompSat}
\end{figure}

In the scenario with impact parameter we find quite different behavior of the solution. As is seen in Fig.~\ref{fig:RunComp1} the evolution of the running coupling (with parent dipole scheme) is actually very fast in the small dipole region, and it is much faster in the large dipole region. This is obvious since  in the large dipole region the coupling is fixed at $\asb = 0.3$ which yields approximately three times as fast an evolution versus the case where $\asb = 0.1$ is fixed.    It can be seen there are box effects beginning to manifest in the running coupling case due to the frozen coupling evolving very quickly in the large dipole regime and reaching the box.

\begin{figure}
\centering
\subfigure[Saturation scale at small dipole size]{\label{fig:RunSat1}\includegraphics[angle=270,width=0.3\textwidth]{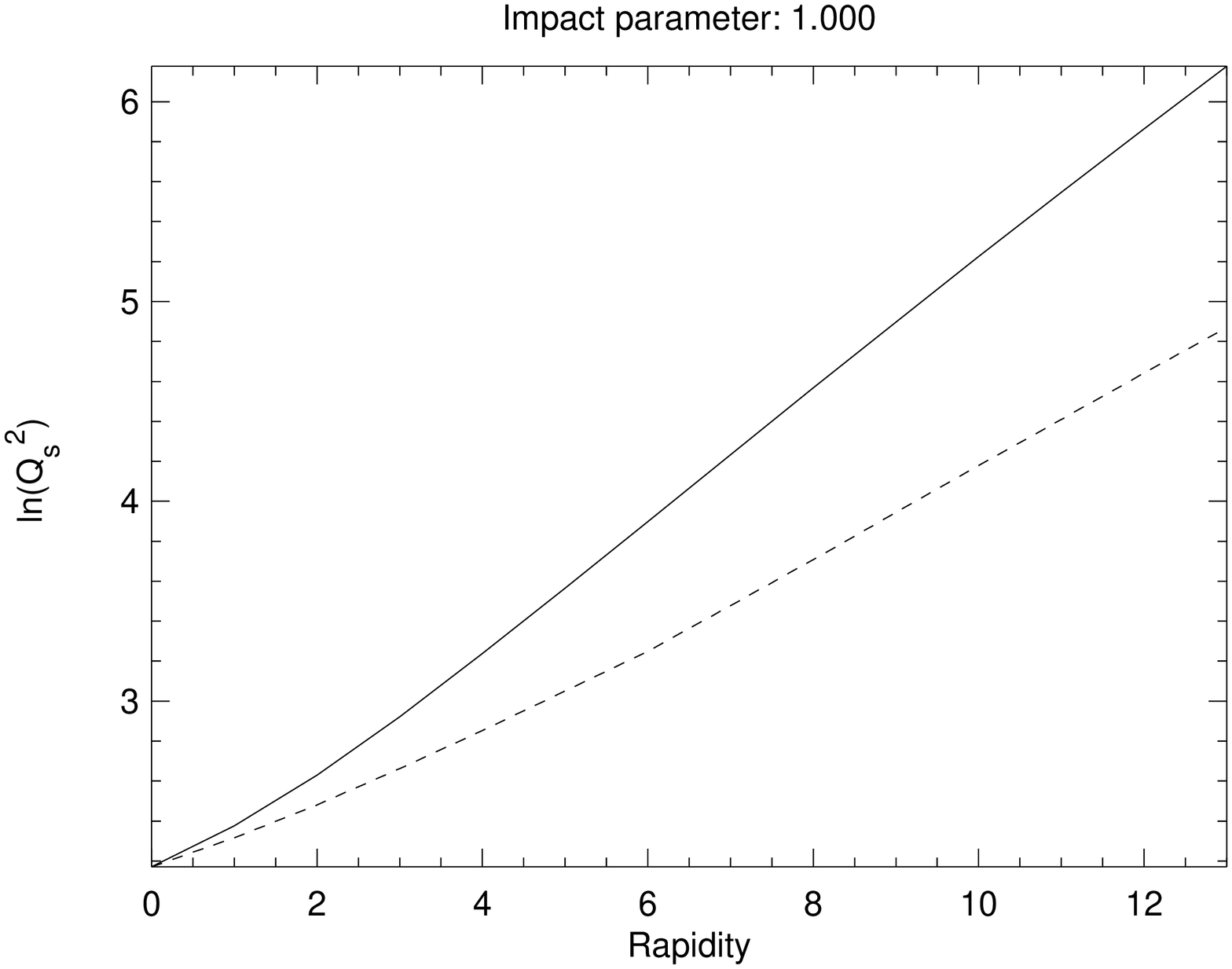}}
\subfigure[Saturation scale at large dipole size]{\label{fig:RunSat2}\includegraphics[angle=270,width=0.3\textwidth]{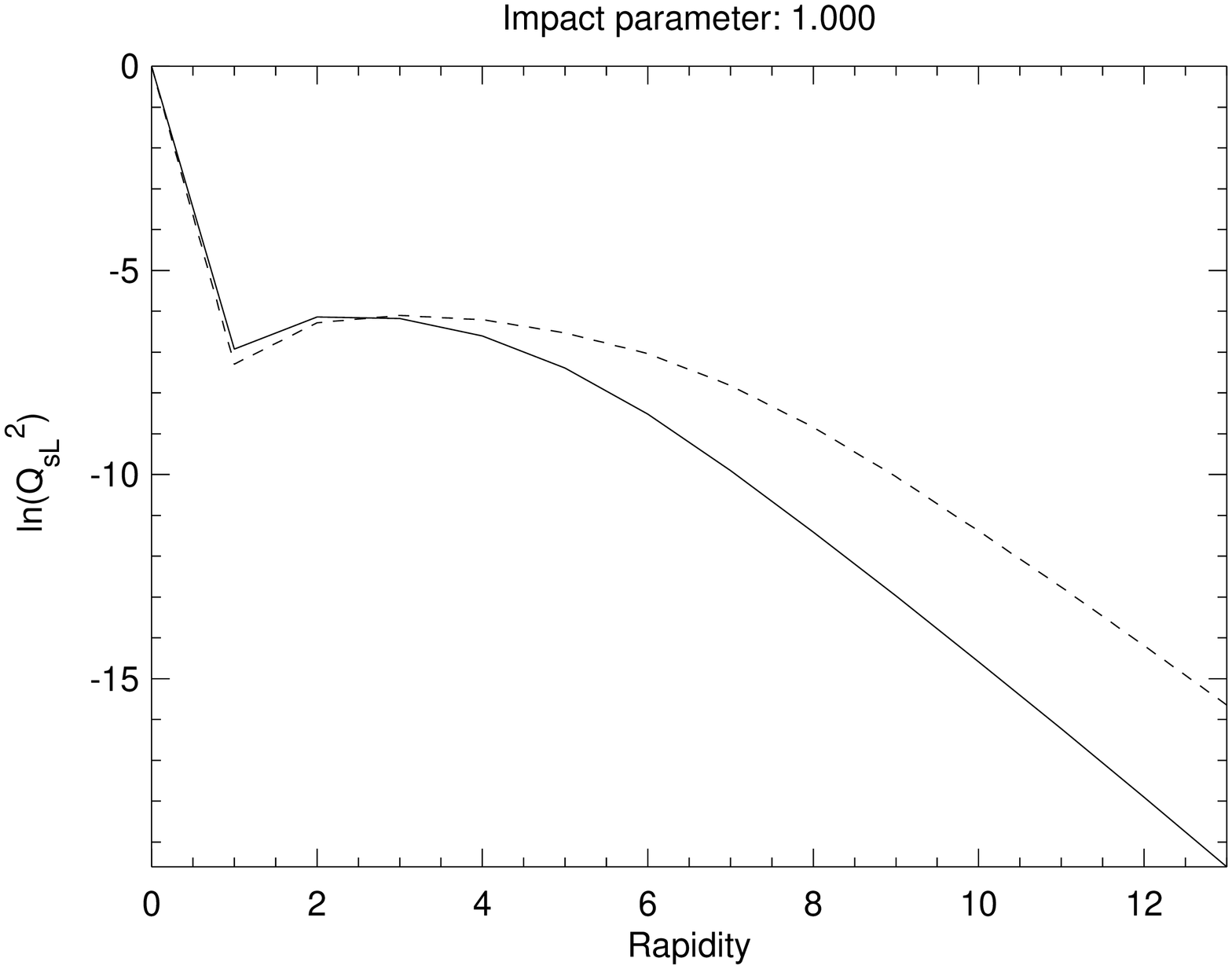}}
\subfigure[Black disc radius in impact parameter]{\label{fig:RunSat3}\includegraphics[angle=270,width=0.3\textwidth]{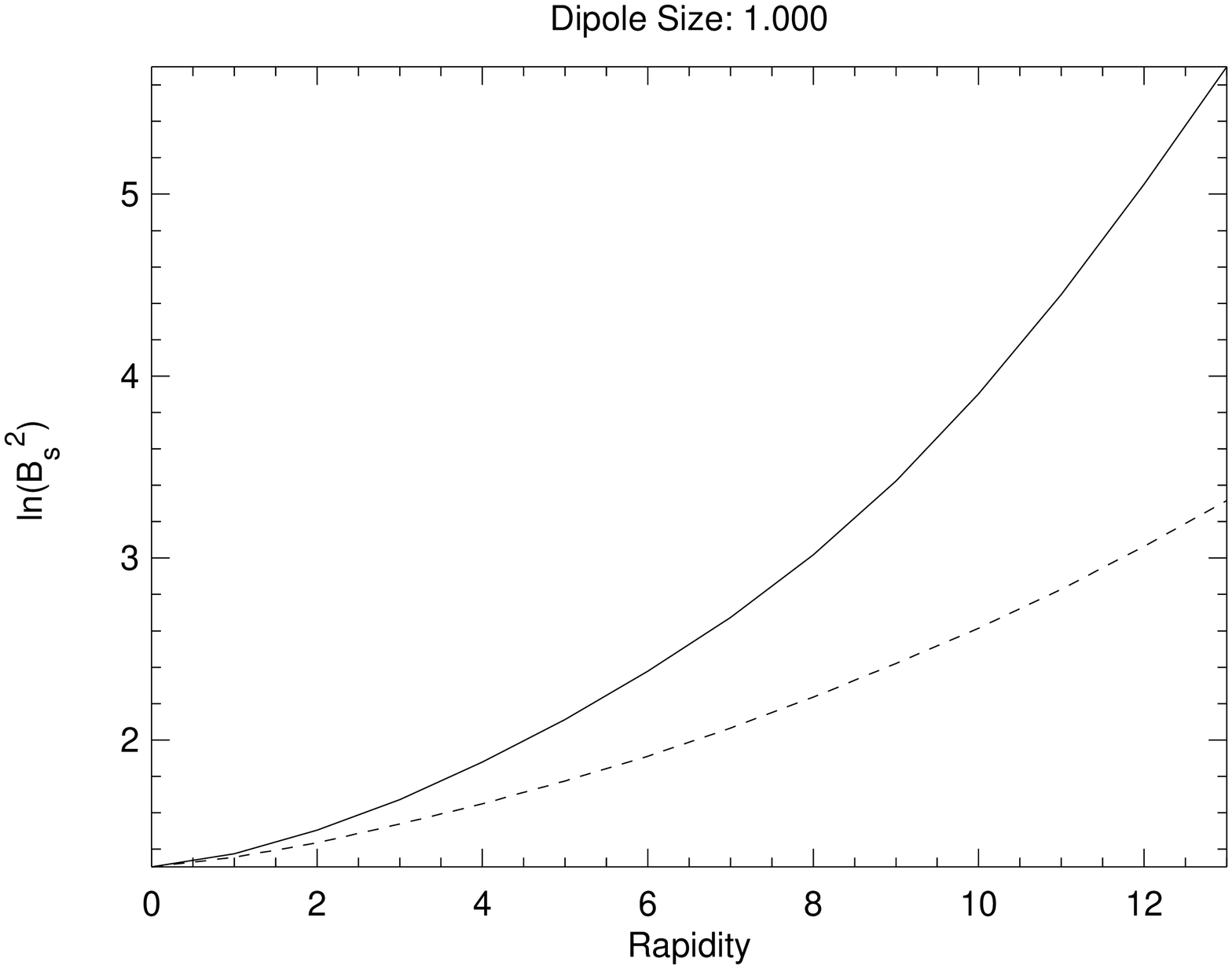}}
\caption{Small dipole saturation scale, large dipole saturation scale and the black disc radius for the case of the running coupling within the parent dipole scheme.
The solid lines are for the LO kernel and dashed lines are for Bessel kernel.}
\label{fig:RunRunCompSat}
\end{figure}

It can be seen in Fig.~\ref{fig:RunRunCompSat} that the  dependence  of the saturation scales on the rapidity is now again almost exponential.  In this case we can extract the exponents  by fitting exponential forms in the rapidity as we did for the fixed coupling case (table \ref{Table:LambdaSummaryRun}).  Note that the definition of the exponents are now different than in the previous section. Here, we took $Q_s \sim \exp(\lambda_s Y), Q_{sL}\sim \exp(-\lambda_{L} Y), B_s\sim \exp(\lambda_B Y)$. The reason that the dependencies are almost exponential is due to the large sensitivity to the infrared and the fact that the coupling is frozen. In that case the solutions behave as almost with the fixed running determined by the freezing value.

Similar pattern is found in the case of the running coupling with the scenario (\ref{eq:Balitsky}).  The only difference is in the small dipole regime where the evolution is slightly slower than
that of the parent dipole scheme.  This can be seen by comparing the extracted exponents in Table \ref{Table:LambdaSummaryRun}.

\begin{center}
\begin{table}
\begin{tabular}{| l || c | c | c | c |}
\hline
 & $\lambda_{s}$ & $\lambda_{L}$ & $\lambda_{B}$ & $\lambda_{BD}$\\
\hline
LO Kernel $\bar{\alpha_s}^{(PD)}$ & 0.30 & 1.68 & 0.60 & 0.65  \\
LO Kernel $\bar{\alpha_s}^{(Bal)}$ & 0.29 & 1.68 & 0.64 & 0.68   \\
Bessel Kernel $\bar{\alpha}_s^{(PD)}$ & 0.22 & 1.42 & 0.24 & 0.32  \\
\hline
\end{tabular}
\caption{Summary of extracted evolution exponents with impact parameter for running coupling case.
${\rm PD}$ means parent dipole prescription, ${\rm Bal}$ means prescription (\ref{eq:Balitsky}). }
\label{Table:LambdaSummaryRun}
\end{table}
\end{center}
\begin{figure}[htb]
\centering
\subfigure[Black disc cross section for the fixed coupling $\alpha_s = 0.1$ case (solid line) versus the running coupling case (dashed line), both for the LO kernel]{\label{fig:RunBDCS1}\includegraphics[angle=270,width=0.42\textwidth]{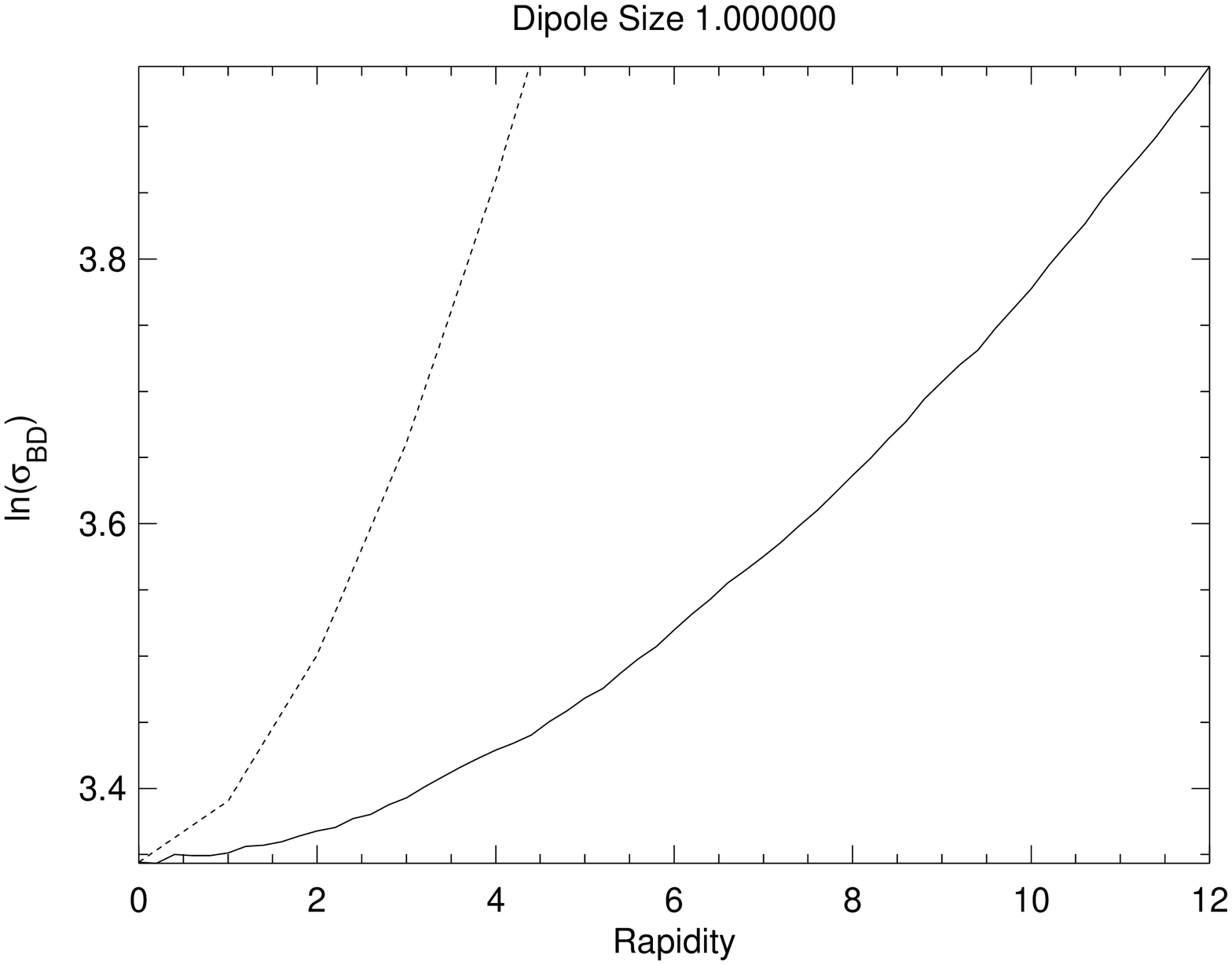}}\hspace*{0.5cm}
\subfigure[Black disc cross section for case of running coupling with the LO kernel (solid line) versus the running coupling case with the Bessel kernel(dashed line)]{\label{fig:RunBDCS2}\includegraphics[angle=270,width=0.42\textwidth]{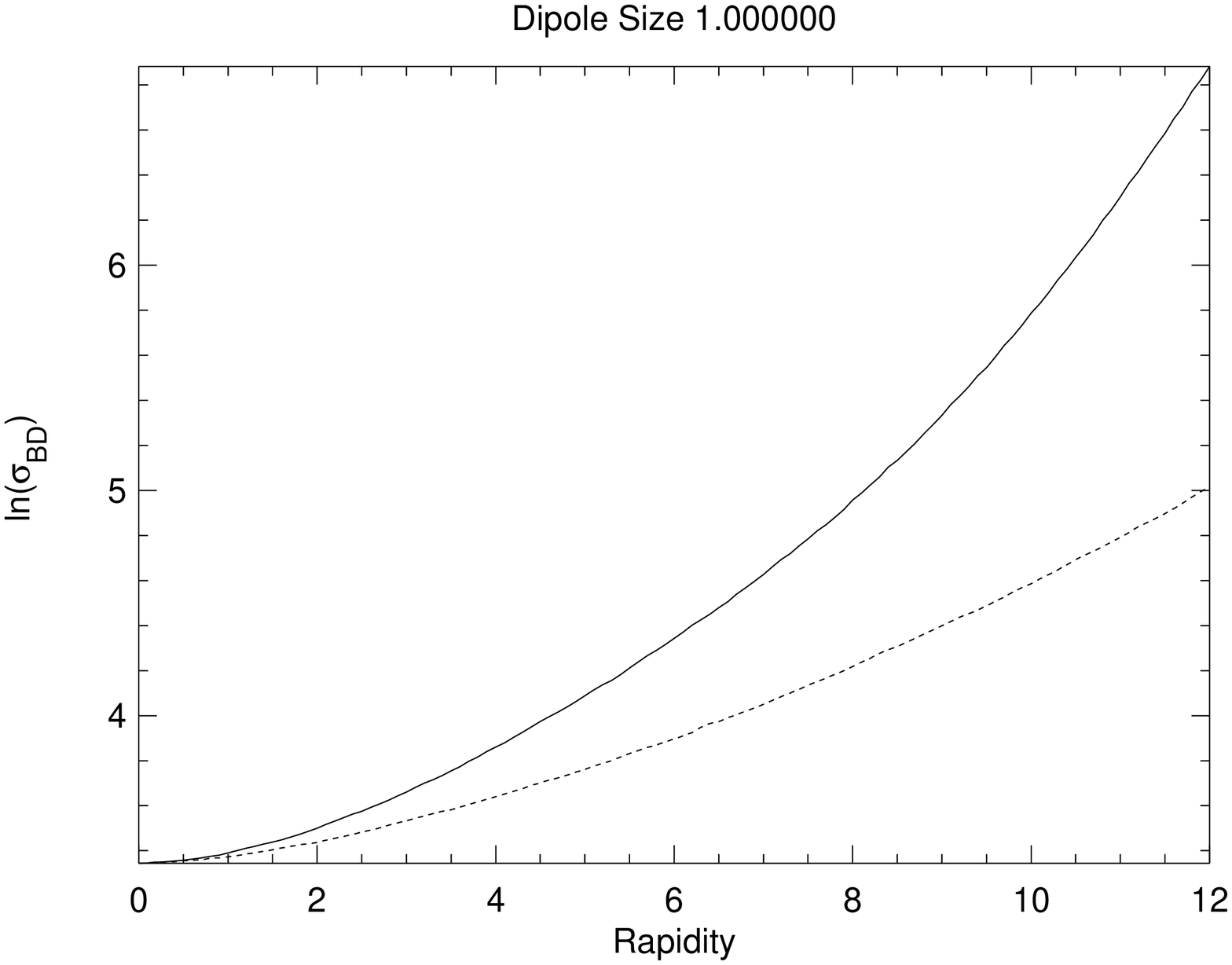}}
\caption{Dipole cross section. Contribution from the 'black disc'.}
\label{fig:RunningBDCS}
\end{figure}

The behavior observed is of course something that has been analyzed before, in the context of the linear BFKL with running coupling \cite{Ciafaloni:2002xk}.  In particular it was observed that, the BFKL solution shows the tunneling scenario, where at some value of rapidity the solution is completely dominated by the infrared region. Strictly speaking we are not observing the tunneling scenario here, due to the fact that we have chosen our initial conditions to be concentrated around rather large dipole sizes where the coupling is already large.  Rather, our solutions are completely dominated by the large coupling values and hence
the saturation scale has nearly exponential dependence on rapidity. It will be interesting to analyze the solution for the initial conditions which are located in the small dipole regime to see if the tunneling occurs here.

We have also evaluated the dipole cross section  coming from the black disc regime in the running coupling scenario,
and we parametrize it in the form
\begin{equation}
\sigma_{BD} = 2 \pi R^2_{BD}(x,Y) = 2 \pi R^{(0)}_{BD}e^{\lambda_{BD} Y} \; .
\end{equation}

The extracted value for the exponent $\lambda_{BD}$ is shown also in Table\ref{Table:LambdaSummaryRun}.   Again the black disc cross section increases very fast due to the 
large value of the coupling in the region of freezing.  We have also compared the solutions in the case of the LO and Bessel kernel, the results are shown in Fig.~\ref{fig:RunningBDCS}. Since the coupling is relatively large, the differences between  the evolution with LO and Bessel kernels are more amplified. 


\section{Conclusions}
\label{sec:conclusions}

Let us summarize the most important points of our analysis.

In the case of the solution with the LO kernel,  the extracted exponents of the saturation scales and the black disc radius are consistent  with values obtained  from  the boundary method. The peaks in impact parameter and in dipole size can be very easily understood from arguments based on the conformal symmetry. The relative strength of the evolution of different saturation scales follows as well from the arguments on the conformal symmetry. In particular the black disc radius has an expansion rate which is twice slower than that of the saturation scale for small dipoles.

For the running coupling scenario, in the case of the solutions with impact parameter we no longer observe the self-regularizing behavior of the nonlinear equation.  This is of course due to the increased sensitivity to the large values of the dipole size. Rather, for the initial conditions chosen
one observes strong dependence on the details of the regularization, and basically the exponents of both   the saturation scales are dominated by the largest value of the coupling. This could be tested in more detail by choosing different initial condition, nevertheless one can expect that for the sufficiently large rapidity,
the solution becomes regularization-sensitive, much like it was observed in earlier simulations.

 The cuts on the large dipole sizes, introduced in the form of the modified kernel have in general very small effect for the case of the impact parameter independent kernel. For the case with the impact parameter they are no longer negligible and reduce the exponent by about  $25\%$  for coupling of $\asb=0.1$. It is important to note that, the modified kernel we have chosen does not account for all the type of kinematical cuts, and therefore other cuts, on the small dipole sizes should be included similarly to what was  done in \cite{Avsar:2005iz}. One could expect therefore even stronger effect in this case.

We therefore conclude that the observed self-regularizing behavior of the local BK equation with the running coupling and  almost complete  insensitivity to the other  NLO corrections appear due to the simplified assumption about the impact parameter independence.

In a broader perspective, it will be interesting to perform the analysis with full NLO kernel or the more correct form of the kinematical cuts, as well as introduce effectively confinement effects. It is also vital to  analyze the impact of the corrections which go beyond the mean field approximation  \cite{Mueller:1996te,Mueller:2004sea,Hatta:2007fg,Mueller:2010fi}.

\section*{Acknowledgments}

We would like to thank Emil Avsar and  Leszek Motyka for interesting discussions. This work was supported  by the  MNiSW grant No. N202 249235  and the DOE OJI grant No. DE - SC0002145.  A.M.S. is supported by the Sloan Foundation.

\bibliographystyle{h-physrev4}
\bibliography{mybib}

\end{document}